\documentclass[english,12pt]{article}
\usepackage[cp1251]{inputenc}
\usepackage{babel}
\usepackage{amsfonts, amsmath}

\newcommand{\be}{\begin{equation}} \newcommand{\ee}{\end{equation}}

\def\A{\ensuremath{\boldsymbol{A}}}

\newcommand{\gsim}{\mathrel{\hbox{\rlap{\lower.55ex\hbox{$\sim$}} \kern-.3em \raise.4ex \hbox{$>$}}}}

\begin{document}
\begin{center}
{\bf Minimal  Quantities  and  Measurability.
 Gravity  in  Measurable Format  and    Natural  Transition  to  High Energies
}\\
\vspace{5mm} Alexander Shalyt-Margolin \footnote{E-mail:
a.shalyt@mail.ru; alexm@hep.by}\\ \vspace{5mm} \textit{Research
Institute for Nuclear Problems,Belarusian State University, 11
Bobruiskaya str., Minsk 220040, Belarus}
\end{center}
 \rm\normalsize \vspace{0.5cm}

\begin{abstract}
A new (more general) definition of the measurability concept not
related to the principle of uncertainty  is given. Then gravity is
studied within the scope of this notion. The measurable format of
General Relativity (GR) is constructed and it is shown that this
format represents its deformation. Passage of the measurable
analog of GR to the high energy (quantum gravitational) region is
considered to show that this is quite natural in the physical sense.
The results obtained are discussed; a further
course of studies by the author is indicated.
\end{abstract}

\section{Introduction}

This paper presents the new results obtained within the scope of
the approach to studies of a quantum theory and gravity in terms
of the {\bf measurability} notion, initiated in
\cite{Shalyt-AHEP2}--\cite{Shalyt-ASTP7}, with the aim to form the
above-mentioned theories proceeding from the variations
(increments) dependent on the existent energies.
\\These theories should not involve the infinitesimal variations
$dt,dx_{i},dp_i,dE,i=1,...,3$ and, in general, any abstract small
quantities $\delta t,\delta x_i,\delta E,\delta p_i,...$.
\\In this paper the results from \cite{Shalyt-ASTP6},\cite{Shalyt-ASTP7}are significantly expanded.
\\ Section 2 gives brief information that is necessary for further consideration.
In Sections 3,4 the author logically develops the basics of the
mathematical apparatus required to construct correctly the gravity
in terms  of {\bf measurable} quantities at low energies $E\ll
E_p$.
\\ Then the results obtained in Sections 3,4, are directly applied to effect
the above-mentioned construction in Section 5. Section 5 is
subdivided into two parts: Subsection 5.1, where the author
derives a {\bf measurable} variant of Einstein Equations
considered as deformation of the corresponding Einstein Equations
in the canonical (classical) theory; and Subsection 5.2, where it
is shown that at low energies $E\ll E_p$  there is a {\bf
measurable} form  of the Least Action Principle in Gravity
generating the above equations. Section 6 demonstrates the ways to
use the obtained results to solve the problems arising at the
junction of the classical gravity and a quantum theory at low
energies $E\ll E_p$. Specifically, the question associated with
involvement of the Closed Time-like Curves (CTC) and of the
Hawking problems for black holes in the theory is touched upon.
\\In Section 7 it is demonstrated that the methods proposed
for low energies $E\ll E_p$ may be in a natural way extended to
the region of high energies $E\approx E_p$. In the process one can
obtain a high-energy (quantum) variant of Einstein Equations in
the {\bf measurable} form.
\\ Finally, Section 8 outlines the problems of prime importance, the solutions
of which are necessary for further studies in this direction,  and
the opportunities offered by the approach.

\section{Measurability. Initial Information in Brief}

Let us briefly consider the earlier results
\cite{Shalyt-AHEP2}--\cite{Shalyt-new1} laying the basis for this
study.
\\It is assumed that there is a minimal (universal) unit for
measurement of the length  $\ell$ corresponding to some maximal
energy $E_\ell=\frac{\hbar c }{\ell}$ and a universal unit for
measurement of time $\tau=\ell/c$. Without loss of generality, we
can consider $\ell$ and $\tau$ at Plank’s level, i.e. $\ell=\kappa
l_p,\tau=\kappa t_p$, where the numerical constant $\kappa$ is on
the order of 1. Consequently, we have $E_\ell\propto E_p$ with the
corresponding   proportionality factor.
\\Then we consider a set of all nonzero momenta
\begin{eqnarray}\label{Meas-D1.G}
{\bf P}=\{p_{x_i}\},i=1,..,3;|p_{x_i}|\neq 0.
\end{eqnarray}
and a subset of the {\bf Primarily Measurable} momenta
characterized by the property
\begin{eqnarray}\label{Meas-D2.G}
p_{x_i}\doteq p_{N_{i}}=\frac{\hbar}{N_{i}\ell},
\end{eqnarray}
where $N_{i}$ is an integer number and $p_{x_i}$ is  the momentum
corresponding to the coordinate $x_i$.
\\Formula (\ref{Meas-D2.G})  gives rise to the following definition:
\\
\\{\bf Definition 1.} {\bf Primary Measurability}
\\{\bf 1.1}. Any variation in $\Delta x_i$ for the coordinates $x_i$
and $\Delta t$ of the time $t$  is considered {\bf primarily
measurable} if
\begin{eqnarray}\label{Meas-D3.}
\Delta x_i=N_{\Delta x_i}\ell,\Delta t=N_{\Delta t}\tau,
\end{eqnarray}
where $N_{\Delta x_i}\neq 0$  and $N_{\Delta t}\neq 0$ are integer
numbers.
\\{\bf 1.2}. Let us define any physical quantity as {\bf
primary or elementary measurable} when its value is consistent
with point  {\bf 1.1} of this Definition.
\\
\\So, from {\bf Definition 1.} it directly follows that all the momenta
satisfying (\ref{Meas-D2.G}) are the {\bf Primarily Measurable}
momenta.
\\Then we consider formula (\ref{Meas-D2.G}) and {\bf Definition 1.}
with the addition of the momenta $p_{x_0}\doteq
p_{N_{0}}=\frac{\hbar}{N_{0}\ell}$, where $N_{0}$ is an integer
number corresponding to the time coordinate ($N_{\Delta t}$ in
formula (\ref{Meas-D3.})).
\\ For convenience, we denote {\bf Primarily Measurable Quantities}
satisfying {\bf Definition 1.} in the abbreviated form as {\bf
PMQ}.
\\ It should be noted,that the space-time quantities
 \begin{eqnarray}\label{Meas-D4}
\frac{\tau}{N_{t}}=p_{N_{t}c}\frac{\ell^{2}}{c\hbar} \nonumber
\\
\frac{\ell}{N_{i}}=p_{N_{i}}\frac{\ell^{2}}{\hbar},1=1,...,3,
\end{eqnarray}
where $p_{N_{i}},p_{N_{t}c}$ are {\bf Primarily Measurable}
momenta, up to the fundamental constants are coincident with
$p_{N_{i}},p_{N_{t}c}$ and they may be involved at any stage of
the calculations but, evidently, they are not  {\bf PMQ} in the
general case.
\\Consequently {\bf PMQ} is inadequate for studies of the physical processes.
Therefore, it is reasonable to use {\bf Definition 2.}
\\
\\{\bf Definition 2.} {\bf Generalized Measurability}
\\We define any physical quantity at all energy scales
as {\bf generalized measurable} or, for simplicity, {\bf
measurable} if any of its values may be obtained in terms of {\bf
PMQ} specified by points {\bf 1.1},{\bf 1.2} of {\bf Definition 1}
\\It is important to make the following remark:
\\{\bf Remark 2.1}
\\As long as $\ell$ is a minimal {\bf measurable} length and
$\tau$ is a minimal {\bf measurable} time, values of all {\it
physical quantities} should agree with this condition, i.e., their
expressions should not involve the lengths $l<\ell$ and the times
$t<\tau$ (and hence the momenta $p>p_\ell$ and the energies
$E>E_\ell$). Because of this, values of the length $\ell/N_{i}$
and of the time $\ell/N_{t}$ from formula (\ref{Meas-D4}) could
not appear in expressions for {\it physical quantities}, being
involved only in intermediate calculations, especially at the
summation for replacement of the infinitesimal quantities
$dt,dx_i;i=1,2,3$ on passage from a continuous theory to its
measurable variant.
\\
\\The main target of the author is to form a quantum theory and gravity
only in terms of {\bf measurable} quantities (or of {\bf PMQ}) in
line with {\bf Remark 2.1}.
\\
\\Now we consider separately the two cases.
\\
\\A) {\bf Low Energies, $E\ll E_{p}$}.
\\ In ${\bf P}$ we consider the domain  ${\it{\bf P}}_{LE}\subset {\bf P}$
(LE is abbreviation of ''Low Energies'') defined by the conditions
\begin{eqnarray}\label{Meas-D1.}
{\it{\bf P}}_{LE}=\{p_{x_i}\},i=1,..,3;P_{\ell}\gg|p_{x_i}|\neq 0,
\end{eqnarray}
where $P_{\ell}=E_{\ell}/c$--maximal momentum.
\\ In this case the formula of (\ref{Meas-D2.G}) takes the form
\begin{eqnarray}\label{Meas-D2.}
N_{i}=\frac{\hbar}{p_{x_i}\ell}, or
\\ \nonumber p_{x_i}\doteq p_{N_{i}}=\frac{\hbar}{N_{i}\ell}
\\ \nonumber|N_{i}|\gg 1,
\end{eqnarray}
where the last row of the formula (\ref{Meas-D2.}) is given by the
requirement (\ref{Meas-D1.}).
\\ As shown in \cite{Shalyt-ASTP6},\cite{Shalyt-ASTP7},
since the energies $E\ll E_{\ell}$ are low, i.e.  ($|N_{i}|\gg 1$),
{\bf primarily measurable} momenta are sufficient to specify the
whole domain of the momenta to a high accuracy ${\it{\bf
P}}_{LE}.$
\\In the indicated domain a
discrete set of {\bf primary measurable} momenta $p_{N_{i}},
(i=1,...,3)$ from  formula (\ref{Meas-D2.}) varies almost
continuously, practically covering the whole domain.
\\That is why further ${\it{\bf P}}_{LE}$ is associated with the domain  of
 {\bf primary measurable} momenta, satisfying the conditions of
the formula (\ref{Meas-D1.}) (or (\ref{Meas-D2.})).
\\ Of course, all the calculations of point A) also comply with the
{\bf primary measurable} momenta $p_{N_{t}c}\doteq p_{N_{0}}$ in
formula (\ref{Meas-D4}). Because of this, in what follows we
understand ${\it{\bf P}}_{LE}$ as a set of the {\bf primary
measurable} momenta $p_{x_{\mu}}=p_{N_{\mu}},(\mu=0,...,3)$ with
$|N_\mu|\gg 1$.
\\
\\{\it It should be noted that,
 as all the experimentally involved energies $E$ are low,
they meet the condition $E\ll E_\ell$, specifically for LHC the
maximal energies are $\approx 10TeV=10^{4}GeV$, that is by 15
orders of magnitude lower than the Planck energy $\approx
10^{19}GeV$. But since the energy $E_\ell$ is on the order of the
Planck energy $E_\ell\propto E_p$, in this case all the numbers
$N_{i}$ for the corresponding momenta will meet the condition
$min|N_{i}|\approx 10^{15}$,i.e., the formula  of
 (\ref{Meas-D2.}). So, all the experimentally involved momenta are
considered to be {\bf primary measurable} momenta,i.e. ${\it{\bf
P}}_{LE}$ at low energies $E\ll E_\ell$}.
\\
\\{\bf Note 2.1}
\\Further for the fixed point $x_\mu$ we use the notion
$p_{x_\mu}=p_{N_{x_\mu}}$  or $p_{x_\mu}=p_{N_{\Delta x_\mu}}$.
\\
\\ Naturally,   the small variation $\Delta p_{x_\mu}$ at the point
$p_{x_\mu}=p_{N_{x_\mu}}$ of the momentum space ${\it{\bf
P}}_{LE}$ is represented by the  {\bf primary measurable} momentum
$p_{N^{'}_{x_\mu}}$ with the property $|N^{'}_{x_\mu}|\gg
|N_{x_\mu}|$.
\\So, in the proposed paradigm at low energies $E\ll
E_p$ a set of the {\bf primarily measurable} ${\it{\bf P}}_{LE}$
is discrete, and in every measurement of  $\mu=0,...,3$ there is
the discrete subset ${\it{\bf P_{x_\mu}}}\subset {\it{\bf
P}}_{LE}$:
\begin{eqnarray}\label{Grav7}
{\it{\bf P_{x_\mu}}}\doteq
\{...,p_{N_{x_\mu}-1},p_{N_{x_\mu}},p_{N_{x_\mu}+1},...\}.
\end{eqnarray}
In this case, as compared to the canonical quantum theory, in
continuous space-time we have the following substitution:
\begin{eqnarray}\label{Div1.4}
dp_{\mu}\mapsto \mathbf{\Delta p_{N_{x_\mu}}=
p_{N_{x_\mu}}-p_{N_{x_\mu}+1}=p_{N_{x_\mu}(N_{x_\mu}+1)}};
 \nonumber\\ \frac{\partial}{\partial p_{\mu}}\mapsto
 \mathbf{\frac{\Delta}{\Delta
p_{\mu}},\frac{\partial F}{\partial p_{\mu}}\mapsto \frac{\Delta
F(p_{N_{x_\mu}})}{\Delta p_{\mu}}
=\frac{F(p_{N_{x_\mu}})-F(p_{N_{x_\mu}+1})}{p_{N_{x_\mu}}-p_{N_{x_\mu}+1}}
=\frac{F(p_{N_{x_\mu}})
-F(p_{N_{x_\mu}+1})}{p_{N_{x_\mu}(N_{x_\mu}+1)}}}.
\end{eqnarray}
It is clear that for sufficiently high integer values of
$|N_{x_\mu}|$, formula (\ref{Div1.4}) reproduces a continuous
paradigm in the momentum space to any preassigned accuracy.
\\Similarly for sufficiently high integer values of $|N_{t}|$ and
$|N_{i}\doteq N_{x_i}|$ , the quantities  $\tau/N_{t},\ell/N_{x_i}$  from
formula (\ref{Meas-D4}) may be arbitrary small.
\\ Hence, for sufficiently high integer values of  $|N_{t}|$
and $|N_{i}\doteq N_{x_i}|$, the quantities
$\tau/N_{t},\ell/N_{x_i}$  are nothing but   a {\bf measurable}
analog  of  the small quantities $\delta x_i,\delta t$ and the
infinitesimal quantities $dx_i,dt$, i.e. $\delta x_\mu$, and
$dx_\mu$, $\mu=0,...,3$.
\\As follows from formula (\ref{Meas-D4}),
for sufficiently high integer values of $|N_{x_\mu}|,\mu=0,...,3$,
the {\bf primarily measurable} momenta ${\it{\bf P_{x_\mu}}}$
(formula (\ref{Grav7})) represent a{\bf measurable} analog of
small (and infinitesimal) space-time {\it increments} in the
space-time variety $\mathcal{M}\subset \mathbf{R^{4}}$.
\\ Because of this, for sufficiently high integer values
of $|N_{x_\mu}|$, the space-time analog of formula (\ref{Div1.4})
is as follows:
\begin{eqnarray}\label{Div1.4new}
dx_{\mu}\mapsto \mathbf{\frac{\ell}{N_{x_\mu}}};
 \nonumber\\ \frac{\partial}{\partial
 x_{\mu}}\mapsto \mathbf{\frac{\Delta}{\Delta_
{N_{x_\mu}}}}, \frac{\partial F}{\partial x_\mu}\mapsto
\mathbf{\frac{\Delta F(x_\mu)}{\Delta_
{N_{x_\mu}}}=\frac{F(x_\mu+\ell/N_{x_\mu})-F(x_\mu)}{\ell/N_{x_\mu}}}.
\end{eqnarray}
{\bf Note 2.2.} In this way any point
$\{x_{\mu}\}\in\mathcal{M}\subset \mathbf{R^{4}}$ and any set of
integer numbers high in absolute values $\{N_{x_\mu}\}$ are
correlated with a system of the neighborhoods for this point
$(x_{\mu}\pm\ell/N_{x_\mu})$. It is clear that, with an increase
in $|N_{x_\mu}|$, the indicated system converges to the point
$\{x_{\mu}\}$. In this case all the ingredients of the initial
(continuous) theory the partial derivatives including are replaced
by the corresponding finite differences.
\\
\\{\bf Principle of Correspondence to Continuous Theory (PCCT).}
\\
\\At low energies $E\ll E_p$ (or same $E\ll E_\ell$)
the infinitesimal space-time quantities $dx_\mu;\mu=0,...,3$ and
also infinitesimal values of the momenta $dp_i,i=1,2,3$ and of the
energies $dE$ form the basic instruments (“construction
materials”) for any theory   in continuous space-time.  Because of
this, to construct the {\bf measurable} variant of such a theory,
we should find the adequate substitutes for these quantities.
\\It is obvious that in the first case the substitute is represented
by the quantities $\ell/N_{x_\mu}$, where $|N_{x_\mu}|$ -- no
arbitrary large (but finite!) integer, whereas in the second case
by
$p_{N_{x_i}}=\frac{\hbar}{N_{x_i}\ell};i=1,2,3;\mathcal{E}_{N_{x_0}}=\frac{c\hbar}{N_{x_0}\ell}$,
where $N_{x_\mu}$ -- integer with the above properties
$\mu=0,...3$.
\\In this way in the proposed approach
all the {\bf primary measurable} momenta
$p_{N_{x_\mu}},|N_{x_\mu}|\gg 1$ are small quantities at low
energies $E\ll E_\ell$ and {\bf primary measurable} momenta
$p_{N_{x_\mu}}$ with  sufficiently large $|N_{{x_\mu}}|\gg 1$
being analogous to {\it infinitesimal} quantities of a continuous
theory.
\\
\\B){\bf High Energies, $E\approx E_{p}$}.
\\ In this case formula (\ref{Meas-D2.G}) takes the form
\begin{eqnarray}\label{Meas-D2.H}
N_{i}=\frac{\hbar}{p_{x_i}\ell}, or
\\ \nonumber p_{x_i}\doteq p_{N_{i}}=\frac{\hbar}{N_{i}\ell}
\\ \nonumber|N_{i}|\approx 1.
\end{eqnarray}
where $N_{i}$ is an integer number and $p_{x_i}$ is  the momentum
corresponding to the coordinate $x_i$. The {\it discrete set}
$p_{N_{i}}\doteq p_{N_{x_i}}$ is introduced as {\bf primarily
measurable} momenta.
\\The main difference of the case B){\bf High Energies}
from the case A) {\bf Low Energies} is in the fact that at {\bf
High Energies} the {\bf primary measurable} momenta are {\it
inadequate} for theoretical studies at the energy scales $E\approx
E_{p}$.
\\
\\Indeed, as it has been shown in \cite{Shalyt-new1},
the Generalized Uncertainty Principle (GUP),which is
 generalization of the Heisenberg Uncertainty Principle (HUP)
\cite{Ven1}--\cite{Nozari}
\begin{equation}\label{U2}
\Delta x\geq\frac{\hbar}{\Delta p}+ \alpha^{\prime}
l_{p}^2\frac{\triangle p}{\hbar},
\end{equation}
where $\alpha^{\prime}$ is a constant on the order of 1, leading
to the minimal length $\ell$ on the order of the Planck length
$\ell\doteq 2 \surd \alpha^{\prime} l_{p}$ at high energies
inevitably results in the momenta $\Delta p(N_{\Delta x}, GUP)$
which are not {\bf primarily measurable}:
\begin{eqnarray}\label{root3.3}
\Delta p\doteq \Delta p(N_{\Delta x},
GUP)=\frac{\hbar}{1/2(N_{\Delta x}+ \sqrt{N_{\Delta
x}^{2}-1})\ell}.
\end{eqnarray}
It is clear that for $N_{\Delta x}\approx 1$ the momentum  $\Delta
p(N_{\Delta x}, GUP)$ is not a {\bf primary measurable} momentum.
\\From {\bf Remark 2.1} in formula (\ref{root3.3}) it follows that
the condition $N_{\Delta x}\geq 2$ should be fulfilled.
\\On the contrary, at low energies $E\ll
E_{p},(E\ll E_\ell)$ the {\bf primary measurable} space quantity
$\Delta x=N_{\Delta x}\ell$, where $N_{\Delta x}\gg 1$ is an
integer number, due to the validity of the limit
\begin{eqnarray}\label{root3.1.} \lim\limits_{N_{\Delta
x}\rightarrow \infty}\sqrt{N_{\Delta x}^{2}-1}=N_{\Delta x},
\end{eqnarray}
leads to the momentum $\Delta p(N_{\Delta x}, HUP)$:
\begin{eqnarray}\label{root3.2.}
\Delta p\doteq \Delta p(N_{\Delta x},
HUP)=\frac{\hbar}{1/2(N_{\Delta x}+ \sqrt{N_{\Delta
x}^{2}-1})\ell}\approx \frac{\hbar}{N_{\Delta
x}\ell}=\frac{\hbar}{\Delta x}.
\end{eqnarray}
It is inferred that, for sufficiently high integer values of
$N_{\Delta x}$ the momentum  $\Delta p(N_{\Delta x}, HUP)$ within
any high accuracy may be considered to be the {\bf primary
measurable} momentum.
\\ Therefore, to study high (Planck’s) energies $E\approx E_{p}$ ,
we need not only {\bf primarily measurable} momenta but also the
{\bf generalized measurable} momenta.
\\
\\{\bf Remark 1.} What is the main point of this Section?
\\
\\{\bf 1.a)} {\it At low energies $E\ll E_{p}$ we replace the abstract
small and  infinitesimal quantities $\delta x_\mu,dx_\mu,\delta
p_\mu,dp_\mu$ incomparable with each other, by the specific small
quantities $\ell/N_{x_{\mu}},p_{N_{x_{\mu}}}$, which may be made
however small at sufficiently high $|N_{x_{\chi}}|$, still being
ordered and comparable. It is very important that the quantities
$\ell/N_{x_{\mu}},p_{N_{x_{\mu}}}$ are directly associated with
the existing energies; for $|N^{'}_{x_{\mu}}|>|N_{x_{\mu}}|$ the
momentum $p_{|N^{'}_{x_{\mu}}|}<p_{|N_{x_{\mu}}|}$ and
$p_{|N^{'}_{x_{\mu}}|}$ corresponds to lower energy than
$p_{|N_{x_{\mu}}|}$ . The same is true for the space variations
$\ell/N^{'}_{x_{\mu}},\ell/N_{x_{\mu}}.$}
\\
\\{\bf 1.b)}{\it At low energies $E\ll E_{p}$
 we should emphasize the difference between
the {\bf primary measurable} momenta $p_{N_{x_{\mu}}}\in {\it{\bf
P}}_{LE}$ and the space-time quantities $\ell/N_{x_{\mu}}$
corresponding to them in accordance with formula (\ref{Meas-D4}).
\\ The first, that is $p_{N_{x_{\mu}}}$ represent
the whole set of the momenta ${\it{\bf P}}_{LE}$ at low energies
$E\ll E_{p}$ in terms of {\bf measurable} quantities, whereas the
second, $\ell/N_{x_{\mu}}$, represent only the {\bf measurable}
small variations of space-time quantities}.
\\
\\{\bf 1.c)} According to {\bf Definition 1.},
in the relativistic case the {\bf primary measurable} energy  is
of the form
\begin{eqnarray}\label{Meas-D2.G-en}
\mathcal{E}=\frac{\hbar c}{N_{0}\ell},N_{0}\doteq N_{x_0},
\end{eqnarray}
where $N_{0}$ is an integer number, and  at low energies $E\ll E_{p}$
it is obvious that $N_{0}\gg 1.$
\\Then at low energies $E\ll E_{p}$ from {\bf Remark 2.2.}
it follows naturally that {\bf primary measurable} energies, to a
high accuracy, cover the whole low-energy spectrum. Then,
considering that the formula
\\
$\mathcal{E}^{2}=\textbf{p}^{2}c^{2}+m^{2}c^{4}$ low energies
$E\ll E_{p}$ \cite{Land1},\cite{Land2} to a high accuracy is valid
in terms of {\bf measurable} quantities and all components of the
vector $\textbf{p}$ are the {\bf primary measurable} momenta, we
can found the mass $m$ in terms of the {\bf measurability} notion
as follows:
\begin{eqnarray}\label{Meas-D2.G-en2}
m^{2}=\frac{\hbar^{2}}{c^{2}}(\frac{1}{N^{2}_{0}\ell^{2}}-\sum_{1\leq
i\leq 3}\frac{1}{N^{2}_{i}\ell^{2}}).
\end{eqnarray}
\\
\\{\bf 1.d)} Finally, it is important to note that actually
the minimal quantities of length $\ell$ and time $\tau$, and also
the maximal quantities of energy $E_\ell$ and momentum
$p_\ell=E_\ell/c$, considered at the beginning of this section,
are minimal and maximal {\bf primarily measurable} quantities,
respectively.

\section{Measurable Metrics and Coordinate Transformations in Gravity}

According to the results from the previous section, the {\bf
measurable} variant of gravity at low energies $E\ll E_p$ should
be formulated in terms of the {\bf measurable} space-time
quantities $\ell/N_{\Delta x_\mu}$ or {\bf primary measurable}
momenta $p_{N_{\Delta x_\mu}}$.
\\Let us consider the case of the random metric $g_{\mu\nu}=g_{\mu\nu}(x)$
\cite{Einst1},\cite{Akhm}, where $x\in R^{4}$ is some point of the
four-dimensional space $R^{4}$ defined in {\bf measurable} terms.
Now, any such point $x\doteq \{x_{\chi}\}\in R^{4}$ and any set of
integer numbers $\{N_{x_{\chi}}\}$ dependent on the point
$\{x_{\chi}\}$ with the property $|N_{x_{\chi}}|\gg 1$ may be
correlated to the ''bundle'' with the base $R^{4}$ as follows:
\begin{eqnarray}\label{GravM.1}
\emph{B}_{N_{x_{\chi}}}\doteq\{x_{\chi},\frac{\ell}{N_{x_{\chi}}}
\}\mapsto\{x_{\chi}\}.
\end{eqnarray}
It is clear that $\lim\limits_{|N_{x_{\chi}}|\rightarrow
\infty}\emph{B}_{N_{x_{\chi}}}= R^{4}.$
\\As distinct from the normal one, the ''bundle'' $\emph{B}_{N_{x_{\chi}}}$
is distinguished only by the fact that the mapping in formula
(\ref{GravM.1}) is not continuous (smooth) but discrete in fibers,
being continuous  in the limit $|N_{x_{\chi}}|\rightarrow \infty$.
\\Then as a {\it canonically measurable prototype}
of the infinitesimal space-time interval square
\cite{Einst1},\cite{Akhm}
\begin{eqnarray}\label{Grav1.4}
ds^2(x) = g_{\mu\nu}(x)dx^\mu dx^\nu
\end{eqnarray}
we take the expression
\begin{eqnarray}\label{Grav1.5}
\Delta s_{N_{x_{\chi}}}^2(x)\doteq
g_{\mu\nu}(x,N_{x_{\chi}})\frac{\ell^{2}}{N_{x_{\mu}}N_{x_\nu}}.
\end{eqnarray}
Here $g_{\mu\nu}(x,N_{x_{\chi}})$ -- metric $g_{\mu\nu}(x)$ from
formula $(\ref{Grav1.4})$ with the property that minimal {\bf
measurable} variation  of metric $g_{\mu\nu}(x)$ in  point $x$ has
form
\begin{eqnarray}\label{Grav1.5d}
\Delta
g_{\mu\nu}(x,N_{x_{\chi}})_{\chi}=g_{\mu\nu}(x+\ell/N_{x_{\chi}},N_{x_{\chi}})-g_{\mu\nu}(x,N_{x_{\chi}}),
\end{eqnarray}
Let  us  denote by $\Delta_{\chi}g_{\mu\nu}(x,N_{x_{\chi}})$
quantity
\begin{eqnarray}\label{Grav1.5d2}
\Delta_{\chi}g_{\mu\nu}(x,N_{x_{\chi}})=\frac{\Delta
g_{\mu\nu}(x,N_{x_{\chi}})_{\chi}}{\ell/N_{x_{\chi}}}.
\end{eqnarray}
It is obvious that in the case under study   the quantity $\Delta
g_{\mu\nu}(x,N_{x_{\chi}})_{\chi}$  is a {\bf measurable} analog
for the infinitesimal increment $dg_{\mu\nu}(x)$ of the $\chi$-th
component $(dg_{\mu\nu}(x))_{\chi}$ in a continuous theory,
whereas the quantity $\Delta_{\chi}g_{\mu\nu}(x,N_{x_{\chi}})$ is
a {\bf measurable} analog of the partial derivative
$\partial_{\chi}g_{\mu\nu}(x)$.
\\ In this manner we obtain the (\ref{GravM.1})-formula induced bundle over the metric manifold $g_{\mu\nu}(x)$:
\begin{eqnarray}\label{GravM.1Bund}
\emph{B}_{g,N_{x_{\chi}}}\doteq g_{\mu\nu}(x,N_{x_{\chi}})\mapsto
g_{\mu\nu}(x).
\end{eqnarray}
Referring to formula (\ref{Meas-D4}), we can see that
(\ref{Grav1.5}) may be written in terms of the {\bf primary
measurable} momenta $(p_{N_{i}},p_{N_{t}})\doteq p_{N_{\mu}}$ as
follows:
\begin{eqnarray}\label{Grav1.5.1}
\Delta s_{N_{x_{\mu}}}^2(x)=
\frac{\ell^{4}}{\hbar^{2}}g_{\mu\nu}(x,N_{x_{\chi}})p_{N_{x_{\mu}}}p_{N_{x_\nu}}.
\end{eqnarray}
Considering that $\ell \propto l_P$ (i.e., $\ell=\kappa l_P$),
where $\kappa=const$ is on the order of 1, in the general case
(\ref{Grav1.5.1}), to within the constant $\ell^{4}/\hbar^{2}$, we
have
\begin{eqnarray}\label{Grav1.5.2}
\Delta s_{N_{x_{\mu}}}^2(x)=
g_{\mu\nu}(x,N_{x_{\chi}})p_{N_{x_{\mu}}}p_{N_{x_\nu}}.
\end{eqnarray}
As follows from the previous formulae, the {\bf measurable}
variant of General Relativity should be defined in the bundle
$\emph{B}_{g,N_{x_{\chi}}}$.
\\Let us consider any coordinate transformation $x^\mu =
x^\mu\left(\bar{x}^\nu\right)$ of the space--time coordinates in
continuous space—time. Then we have
\begin{eqnarray}\label{Grav2.1}
 dx^\mu = \frac{\partial
x^\mu}{\partial \bar{x}^\nu} \, d\bar{x}^\nu.
\end{eqnarray}
As mentioned at the beginning of this section, in terms of {\bf
measurable} quantities we have the substitution
\begin{eqnarray}\label{Grav2.1change}
 dx^\mu \mapsto \frac{\ell}{N_{\Delta x_\mu}}; d\bar{x}^\nu\mapsto
 \frac{\ell}{\bar{N}_{\Delta
\bar{x}_\nu}},
\end{eqnarray}
where  $N_{\Delta x_\mu},\bar{N}_{\Delta \bar{x}_\nu}$ -- integers
($|N_{\Delta x_\mu}|\gg 1,|\bar{N}_{\Delta \bar{x}_\nu}|\gg 1$)
sufficiently high in absolute value, and hence in the {\bf
measurable} case (\ref{Grav2.1}) is replaced by
\begin{eqnarray}\label{Grav2.1M}
\frac{\ell}{N_{\Delta x_\mu}} =
\Delta_{\mu\nu}(x^\mu,\bar{x}^\nu,1/N_{\Delta
x_\mu},1/\bar{N}_{\Delta \bar{x}_\nu}) \,
\frac{\ell}{\bar{N}_{\Delta \bar{x}_\nu}}.
\end{eqnarray}
Equivalently, in terms of the {\bf primary measurable} momenta we
have
\begin{eqnarray}\label{Grav2.1M-M}
p_{N_{\Delta x_\mu}} =
\Delta_{\mu\nu}(x^\mu,\bar{x}^\nu,1/N_{\Delta
x_\mu},1/\bar{N}_{\Delta \bar{x}_\nu}) \, p_{\bar{N}_{\Delta
\bar{x}_\nu}},
\end{eqnarray}
where $\Delta_{\mu\nu}(x^\mu,\bar{x}^\nu,1/N_{\Delta
x_\mu},1/\bar{N}_{\Delta \bar{x}_\nu})\doteq
\Delta_{\mu\nu}(x^\mu,\bar{x}^\nu,p_{N_{\Delta
x_\mu}},p_{\bar{N}_{\Delta \bar{x}_\nu}})$ -- corresponding matrix
represented in terms of {\bf measurable} quantities.
\\It is clear that, in accordance with formula (\ref{Meas-D4}),
in passage to the limit we get
\begin{eqnarray}\label{Grav2.1M-lim}
\lim\limits_{|N_{\Delta x_\mu}|\rightarrow
\infty}\frac{\ell}{N_{\Delta x_\mu}} = dx^\mu = \nonumber \\=
\lim\limits_{|\bar{N}_{\Delta \bar{x}_\nu}|\rightarrow
\infty}\Delta_{\mu\nu}(x^\mu,\bar{x}^\nu,1/N_{\Delta
x_\mu},1/\bar{N}_{\Delta \bar{x}_\nu}) \,
\frac{\ell}{\bar{N}_{\Delta \bar{x}_\nu}}= \frac{\partial
\bar{x}^\mu}{\partial x^\nu} \,dx^\nu.
\end{eqnarray}
Equivalently, passage to the limit (\ref{Grav2.1M-lim}) may be
written in terms of  the {\bf primary measurable} momenta
$p_{N_{\Delta x_\mu}},p_{\bar{N}_{\Delta \bar{x}_\nu}}$ multiplied
by the constant  $\ell^{2}/\hbar.$
\\ How we understand formulae (\ref{Grav2.1change})--(\ref{Grav2.1M-lim})?
\\The initial (continuous) coordinate transformations $x^\mu =
x^\mu\left(\bar{x}^\nu\right)$ gives the matrix $\frac{\partial
x^\mu}{\partial \bar{x}^\nu}$. Then, for the integers sufficiently
high in absolute value $\bar{N}_{\Delta
\bar{x}_\nu},|\bar{N}_{\Delta \bar{x}_\nu}|\gg 1$, we can derive
\begin{eqnarray}\label{Grav2.1M-new}
\frac{\ell}{N_{\Delta x_\mu}} = \frac{\partial x^\mu}{\partial
\bar{x}^\nu} \, \frac{\ell}{\bar{N}_{\Delta \bar{x}_\nu}},
\end{eqnarray}
where $|N_{\Delta x_\mu}|\gg 1$ but the numbers for $N_{\Delta
x_\mu}$ are not necessarily integer. However it is easy to see
 the difference between $\ell/N_{\Delta x_\mu}$ and
$\ell/[N_{\Delta x_\mu}]$ (and hence between $p_{N_{\Delta
x_\mu}}$ and $p_{[N_{\Delta x_\mu}]}$) is negligible.
\\ Then substitution of $[N_{\Delta x_\mu}]$  for $N_{\Delta x_\mu}$
in the left-hand side of (\ref{Grav2.1M-new}) leads to replacement
of the initial matrix $\frac{\partial x^\mu}{\partial
\bar{x}^\nu}$ by the matrix
$\Delta_{\mu\nu}(x^\mu,\bar{x}^\nu,1/N_{\Delta
x_\mu},1/\bar{N}_{\Delta \bar{x}_\nu})$ represented  in terms of
{\bf measurable} quantities and, finally, to the formula
(\ref{Grav2.1M}). Clearly, for sufficiently high $|N_{\Delta
x_\mu}|,|\bar{N}_{\Delta \bar{x}_\nu}|$ , the matrix
$\Delta_{\mu\nu}(x^\mu,\bar{x}^\nu,1/N_{\Delta
x_\mu},1/\bar{N}_{\Delta \bar{x}_\nu})$  may be selected no matter
how close to $\frac{\partial x^\mu}{\partial \bar{x}^\nu}$.
\\ Similarly, in the {\bf measurable} format we can get the formula
\begin{eqnarray}\label{Grav2.1new}
 d\bar{x}^\mu = \frac{\partial
\bar{x}^\mu}{\partial x^\nu} \,dx^\nu.
\end{eqnarray}
and correspondingly the matrix
$\widetilde{\Delta_{\mu\nu}}(x^\mu,\bar{x}^\nu,1/N_{\Delta
x_\mu},1/\bar{N}_{\Delta \bar{x}_\nu})$ instead of the matrix
$\Delta_{\mu\nu}(x^\mu,\bar{x}^\nu,1/N_{\Delta
x_\mu},1/\bar{N}_{\Delta \bar{x}_\nu})$.
\\Thus, any coordinate
transformations may be represented, to however high accuracy, by
the {\bf measurable} transformation (i.e., written in terms of
{\bf measurable} quantities), where the principal components are
the {\bf measurable} quantities $\ell/N_{\Delta x_\mu}$  or the
{\bf primary measurable} momenta $p_{N_{\Delta x_\mu}}$.

\section{Measurability and Its Mathematical Instruments in Gravity}

Using the results from the previous section, we can define the {\bf
measurable} analogs of the principal ingredients of General Relativity.
In the canonical (continuous) theory the vector $A^\mu$ is referred
to as \underline{contravariant} if it transforms, under the
coordinate transformation, in the same way as coordinates do:
\begin{eqnarray}\label{Cont1}
A^\mu(x) = \frac{\partial x^\mu}{\partial \bar{x}^\nu} \,
\bar{A}^\nu\left(\bar{x}\right).
\end{eqnarray}
In the {\bf measurable} case the formula for (\ref{Cont1}) is as follows:
\begin{eqnarray}\label{Meas1}
 A^\mu(x,{N_{\Delta x_{\chi}}}) =
\Delta_{\mu\nu}(x^\mu,\bar{x}^\nu,1/N_{\Delta
x_\mu},1/\bar{N}_{\Delta \bar{x}_\nu}) \ \,
\bar{A}^\nu\left(\bar{x},\bar{N}_{\Delta x_{\chi}}\right).
\end{eqnarray}
Here the \underline{contravariant} vector $A^\mu(x,{N_{\Delta
x_{\chi}}})$, by virtue of replacement of the matrix  $\frac{\partial
x^\mu}{\partial \bar{x}^\nu}$ by
$\Delta_{\mu\nu}(x^\mu,\bar{x}^\nu,1/N_{\Delta
x_\mu},1/\bar{N}_{\Delta \bar{x}_\nu})$, is dependent not only on the point
 $x$ but also on the integer numbers  $N_{\Delta x_{\chi}},|N_{\Delta
x_{\chi}}|\gg 1$ (i.e. on minimal variations in $\ell/N_{\Delta x_{\chi}}$  at the point $x$).
\\ Similarly, in terms of $A_\mu(x,{N_{\Delta
x_{\chi}}})$, we can define a   {\bf measurable} analog of the
\underline{covariant} vector $A_\mu(x)$ with the transformational
properties
\begin{eqnarray}\label{Meas2}
 A^\mu(x,{N_{\Delta x_{\chi}}}) =
\widetilde{\Delta_{\mu\nu}}(x^\mu,\bar{x}^\nu,1/N_{\Delta
x_\mu},1/\bar{N}_{\Delta \bar{x}_\nu}) \ \,
\bar{A}^\nu\left(\bar{x},\bar{N}_{\Delta x_{\chi}}\right).
\end{eqnarray}
Let us consider in a continuous theory the equality of the two infinitesimal intervals
\begin{eqnarray}\label{Cont3.1}
ds^2 = g_{\mu\nu}(x) \, dx^\mu dx^\nu =
\bar{g}_{\alpha\beta}\left(\bar{x}\right) \, d\bar{x}^\alpha
d\bar{x}^\beta \quad,{\rm or} \nonumber \\
 \quad g_{\mu\nu}(x) =
\frac{\partial \bar{x}^\alpha}{\partial x^\mu} \, \frac{\partial
\bar{x}^\beta}{\partial x^\nu}\, \bar{g}_{\alpha \beta}
\left(\bar{x}\right).
\end{eqnarray}
Obviously, the {\bf measurable} analog of the upper line of
formula (\ref{Cont3.1}) should be the equality
\begin{eqnarray}\label{Meas3.1}
\Delta s^2\approx g_{\mu\nu}(x,N_{\Delta
x^{\chi}})\frac{\ell^{2}}{N_{\Delta x^{\mu}}N_{\Delta
x^\nu}}\approx\bar{g}_{\alpha\beta}(\bar{x},\bar{N}_{\Delta
\bar{x}^{\chi}})\frac{\ell^{2}}{\bar{N}_{\Delta
\bar{x}^{\alpha}}\bar{N}_{\Delta \bar{x}^\beta}}.
\end{eqnarray}
In formula (\ref{Meas3.1})and in all subsequent formulae the
approximation sign $\approx$ everywhere denotes however high
accuracy for sufficiently great $|N_{\Delta x^{\mu}}|,|N_{\Delta
x^{\nu}}|,|\bar{N}_{\Delta \bar{x}^{\alpha}}|,|\bar{N}_{\Delta
\bar{x}^{\beta}}|$.
\\ Nevertheless, in the case of formula (\ref{Cont3.1}) we involve
infinitesimal intervals, whereas in the case of formula
(\ref{Meas3.1}) we consider the intervals having {\it small} but
finite value. Because of this, it is assumed that formula
(\ref{Meas3.1}) is valid when the second and the third terms in
this formula have the same sign (or both are zero), so
\begin{eqnarray}\label{Meas3.2}
g_{\mu\nu}(x,N_{\Delta x^{\chi}})\frac{\ell^{2}}{N_{\Delta
x^{\mu}}N_{\Delta
x^\nu}}-\bar{g}_{\alpha\beta}(\bar{x},\bar{N}_{\Delta
\bar{x}^{\chi}})\frac{\ell^{2}}{\bar{N}_{\Delta
\bar{x}^{\alpha}}\bar{N}_{\Delta \bar{x}^\beta}}\approx 0.
\end{eqnarray}
Then from the foregoing results it follows that a {\bf measurable}
analog of the lower line in formula (\ref{Cont3.1}) takes the form
\begin{eqnarray}\label{Meas3.2}
g_{\mu\nu}(x,N_{\Delta x^{\chi}})\approx\nonumber
\\\approx
\widetilde{\Delta_{\mu\alpha}}(x^\mu,\bar{x}^\alpha,1/N_{\Delta
x_\mu},1/\bar{N}_{\Delta
\bar{x}_\alpha})\widetilde{\Delta_{\nu\beta}}(x^\nu,\bar{x}^\beta,1/N_{\Delta
x_\nu},1/\bar{N}_{\Delta
\bar{x}_\beta})\bar{g}_{\alpha\beta}(\bar{x},\bar{N}_{\Delta
\bar{x}^{\chi}}).
\end{eqnarray}
{\bf Remark 4.1}
\\It is clear that for sufficiently great
$|N_{\Delta x_{\chi}}|,|\bar{N}_{\Delta \bar{x}_{\chi}}|$ all the
above given formulae in a {\bf measurable} variant may be no
matter how close to the corresponding formulae associated with a
continuous consideration. Without loss of generality, this may be
assumed for lowering or growing indices of the contravariant or
covariant vectors, tensors, etc.
\\ Specifically, in a {\bf measurable} variant we have
\begin{eqnarray}\label{Meas4.1}
A_\mu(x,{N_{\Delta x_{\chi}}}) = g_{\mu\nu}(x,N_{\Delta
x^{\chi}})A^\nu(x,{N_{\Delta x_{\chi}}})\approx g_{\mu\nu} \,
A^\nu(x,{N_{\Delta x_{\chi}}}),\nonumber
\\A^{\mu}(x,{N_{\Delta x_{\chi}}}) = g^{\mu\nu}(x,N_{\Delta
x^{\chi}})A_\nu(x,{N_{\Delta x_{\chi}}})\approx g^{\mu\nu} \,
A_\nu(x,{N_{\Delta x_{\chi}}})
 \quad \nonumber
\\{\rm and} \quad g^{\mu\nu}(x,N_{\Delta
x^{\chi}}) \,g_{\nu\alpha}(x,N_{\Delta x^{\chi}})\approx
g^{\mu\nu} \, g_{\nu\alpha} = \delta^\mu_\alpha.
\end{eqnarray}
Now we consider formula (\ref{Cont3.1}) in the assumption that the
coordinates $\bar{x}$ belong to a flat space, i.e. the metric
$\bar{g}_{\alpha\beta}(\bar{x})$  is the Lorentzian metric
$\bar{g}_{\alpha\beta}(\bar{x})=\eta_{\alpha\beta}(\bar{x})$:
\begin{eqnarray}\label{Cont4.1}
||\eta_{\alpha\beta}|| = ||\eta^{\alpha\beta}||
=Diag\left(-1,1,1,1\right).
\end{eqnarray}
In this case formula (\ref{Meas3.2}) takes the form
\begin{eqnarray}\label{Meas3.2M}
g_{\mu\nu}(x,N_{\Delta x^{\chi}})=\nonumber
\\=
\widetilde{\Delta_{\mu\alpha}}(x^\mu,\bar{x}^\alpha,1/N_{\Delta
x_\mu},1/\bar{N}_{\Delta
\bar{x}_\alpha})\widetilde{\Delta_{\nu\beta}}(x^\nu,\bar{x}^\beta,1/N_{\Delta
x_\nu},1/\bar{N}_{\Delta
\bar{x}_\beta})\eta_{\alpha\beta}(\bar{x},\bar{N}_{\Delta
\bar{x}^{\chi}}).
\end{eqnarray}
Taking the determinant for both parts (\ref{Meas3.2M}) and assuming that,
by virtue of {\bf Remark 4.1},
$det||\eta_{\alpha\beta}(\bar{x},\bar{N}_{\Delta
\bar{x}^{\chi}}||\approx det||\eta_{\alpha\beta}||=-1$,
we obtain
\begin{eqnarray}\label{Meas3.2M2}
det||g_{\mu\nu}(x,N_{\Delta x^{\chi}})||\approx\nonumber
\\\approx
-J^{2}[{\widetilde{\Delta_{\mu\alpha}}(x^\mu,\bar{x}^\alpha,1/N_{\Delta
x_\mu},1/\bar{N}_{\Delta \bar{x}_\alpha})}],
\end{eqnarray}
where
$J[{\widetilde{\Delta_{\mu\alpha}}(x^\mu,\bar{x}^\alpha,1/N_{\Delta
x_\mu},1/\bar{N}_{\Delta \bar{x}_\alpha})}]$ -- {\bf measurable}
variant of the Jacobian $J$ for the matrix $\frac{\partial
x^\mu}{\partial \bar{x}^\nu}$ when coordinates of the point
$\bar{x}$ belong to a flat space. In view of the lower line in
formula
 (\ref{Meas4.1}), we have \\$det||g_{\mu\nu}(x,N_{\Delta
x^{\chi}})||=1/det||g^{\mu\nu}(x,N_{\Delta x^{\chi}})||$. If,
according to a continuous theory, we introduce the notation
$det||g^{\mu\nu}(x,N_{\Delta x^{\chi}})||\doteq g(N_{\Delta
x^{\chi}})$, the formula (\ref{Meas3.2M2}) should be rewritten as
\begin{eqnarray}\label{Meas3.2M3}
\frac{1}{g(N_{\Delta x^{\chi}})}\approx\nonumber
\\\approx
-J^{2}[{\widetilde{\Delta_{\mu\alpha}}(x^\mu,\bar{x}^\alpha,1/N_{\Delta
x_\mu},1/\bar{N}_{\Delta \bar{x}_\alpha})}],\nonumber
\\{\rm or} \quad  J[{\widetilde{\Delta_{\mu\alpha}}(x^\mu,\bar{x}^\alpha,1/N_{\Delta
x_\mu},1/\bar{N}_{\Delta
\bar{x}_\alpha})}]\approx\frac{1}{\sqrt{-g(N_{\Delta x^{\chi}})}},
\end{eqnarray}
in a complete agreement with the well-known relation in a continuous theory
$J=1/\sqrt{-g}$   for $g\doteq det||g^{\mu\nu}||$
\cite{Land1}.
\\Let $d\Omega$ be the element
of integration with respect to the four-dimensional volume
$\Omega$ in the continuous case. We denote by $\Delta_{(N_{\Delta
x_{\mu}})}\Omega$ its  {\bf measurable} analog.
$\Delta_{(N_{\Delta x_{\mu}})}\Omega$  may be obtained by
substitution of $dx_\mu\mapsto\ell/N_{\Delta x_{\mu}}$ in the
formula for $d\Omega$.
\\ Obviously, we have
\begin{eqnarray}\label{Div1.6}
\Delta_{(N_{\Delta x_{\mu}})}\Omega\equiv
\frac{\ell^{8}}{\hbar^{4}}\Delta_{(p_{N_{\Delta x_{\mu}}})}\Omega.
\end{eqnarray}
The transition from the infinitesimal volume in a flat space
$d\bar{\Omega}$ to the infinitesimal volume in a curved space
$d\Omega$  with the metric $g_{\mu\nu}$, that in the continuous
case takes the form
\begin{eqnarray}\label{Trans-C}
d\bar{\Omega}\mapsto\sqrt{-g}d\Omega \quad or \quad equivalently
\quad d\bar{\Omega}\mapsto\sqrt{|g|}d\Omega,
\end{eqnarray}
in the {\bf measurable} case  is as follows:
\begin{eqnarray}\label{Trans-M}
\Delta_{(\bar{N}_{\Delta
\bar{x}_{\mu}})}\bar{\Omega}\mapsto\sqrt{-g(N_{\Delta
x^{\chi}})}\Delta_{(N_{\Delta x_{\mu}})}\Omega   \quad or \quad
equivalently \quad  \nonumber
\\\Delta_{(\bar{N}_{\Delta
\bar{x}_{\mu}})}\bar{\Omega}\mapsto\sqrt{|g(N_{\Delta
x^{\chi}})|}\Delta_{(N_{\Delta x_{\mu}})}\Omega.
\end{eqnarray}
In a similar way we can obtain other quantities in the
four-dimensional ''{\bf measurable}'' geometry as well. In
particular, the infinitesimal element of the two-dimension surface
in a continuous theory
\begin{eqnarray}\label{Cont5}
df^{ik}=dx^{i}dx'^{k}-dx^{k}dx'^{i}
\end{eqnarray}
in a {\bf measurable} variant  is associated with the element
\begin{eqnarray}\label{Meas5}
\Delta f^{ik}(N_{\Delta x},N_{\Delta x'})=\frac{\ell^2}{N_{\Delta
x^{i}}N_{\Delta x'^{k}}} -\frac{\ell^2}{N_{\Delta x^{k}}N_{\Delta
x'^{i}}}.
\end{eqnarray}
And the infinitesimal three-dimensional volume element
\begin{equation}\label{Cont6}
dS^{ikl}=det\begin{pmatrix}dx^i&dx'^i&dx''^i\nonumber
\\
dx^k&dx'^k&dx''^k\nonumber
\\
dx^l&dx'^l&dx''^l\end{pmatrix}
\end{equation}
is associated with the {\bf measurable}  element
\begin{equation}\label{Meas6}
\Delta S^{ikl}(N_{\Delta x},N_{\Delta x'},N_{\Delta
x''})=det\begin{pmatrix} \frac{\ell}{N_{\Delta
x^{i}}}&\frac{\ell}{N_{\Delta x'^{i}}}&\frac{\ell}{N_{\Delta
x''^{i}}}\nonumber
\\
\frac{\ell}{N_{\Delta x^{k}}}&\frac{k}{N_{\Delta
x'^{k}}}&\frac{\ell}{N_{\Delta x''^{k}}}\nonumber
\\
\frac{\ell}{N_{\Delta x^{l}}}&\frac{\ell}{N_{\Delta
x'^{l}}}&\frac{\ell}{N_{\Delta x''^{l}}}
\end{pmatrix}
\end{equation}
retaining, for sufficiently high $|N_{\Delta x^{i}}|,|N_{\Delta
x^{k}}|,|N_{\Delta x^{l}}|$, all basic relations of a continuous
geometry  (paragraphs 6,83 in \cite{Land1}) to however high
accuracy.

\section{Gravity in Measurable Form at Low Energies}

\subsection{Einstein Equations in Measurable Format as Deformation
of Canonical Theory}
As directly follows from previous results,
specifically from formulae
(\ref{Grav1.5})--(\ref{Grav1.5d2}),formulae of Sections 3,4, the
principal components involved in gravitational equations of
General Relativity have {\bf measurable} form.
\\In particular, the {\it Christoffel symbols} \cite{Einst1},\cite{Akhm}
\begin{eqnarray}\label{Grav1.5.3}
\Gamma^\alpha_{\mu\nu}(x) = \frac12 \, g^{\alpha\beta}(x)\,
\left(\partial_\nu^{\phantom{\frac12}} g_{\beta\mu}(x) +
\partial_\mu \, g_{\nu\beta}(x)-\partial_\beta \,
g_{\mu\nu}(x)\right)
\end{eqnarray}
have the {\bf measurable} format \cite{Shalyt-ASTP6}
\begin{eqnarray}\label{Grav1.5.3M}
\Gamma^\alpha_{\mu\nu}(x,N_{\Delta x_{\chi}}) = \frac12 \,
g^{\alpha\beta}(x,N_{\Delta x_{\chi}}) \, (\Delta_{\nu}
g_{\beta\mu}(x,N_{\Delta x_{\chi}}) + \Delta_{\mu}
g_{\nu\beta}(x,N_{\Delta x_{\chi}})- \nonumber \\
-\Delta_{\beta}g_{\mu\nu}(x,N_{\Delta x_{\chi}})).
\end{eqnarray}
In is simple to derive {\bf measurable} analogs of the well-known
quantities in a continuous theory arising in the definition of the
parallel translation.
\\ Specifically, the formula
\begin{eqnarray}\label{Grav1.5.Cont}
\delta A_\mu = \Gamma^\nu_{\mu\alpha}(x) \, A_\nu(x) \, dx^\alpha,
\end{eqnarray}
in a {\bf measurable}  variant is of the form
\begin{eqnarray}\label{Grav1.5.Meas}
\delta_{meas}A_\mu(x,N_{\Delta x_{\chi}}) =
\Gamma^\nu_{\mu\alpha}(x,N_{\Delta x_{\chi}}) \, A_\nu(x,N_{\Delta
x_{\chi}}) \, \frac{\ell}{N_{\Delta x_{\alpha}}}.
\end{eqnarray}
Note that a formula for the covariant derivative
\begin{eqnarray}\label{Grav1.6.Cont}
A^\mu_{;\alpha} \equiv D_\alpha A^\mu =
\partial_\alpha A^\mu + \Gamma^\mu_{\nu\alpha} \,
 A^\nu = \left(\partial_\alpha \, \delta^\mu_\nu + \Gamma^\mu_{\nu\alpha}\right) \, A^\nu, \nonumber \\
A_{\mu;\, \alpha} \equiv D_\alpha A_\mu = \partial_\alpha A_\mu -
\Gamma^\nu_{\mu\alpha} \, A_\nu = \left(\partial_\alpha \,
\delta^\nu_\mu - \Gamma^\nu_{\mu\alpha}\right)\, A_\nu
\end{eqnarray}
in a {\bf measurable}  format is given as
\begin{eqnarray}\label{Grav1.6.Meas}
\widetilde{A}^\mu_{;\alpha} \equiv \widetilde{D}_\alpha
A^\mu(x,N_{\Delta x_{\chi}})= \mathbf{\frac{\Delta}{\Delta_
{N_{\Delta x_\alpha}}}} A^\mu(x,N_{\Delta x_{\chi}})+
\Gamma^\mu_{\nu\alpha}(x,N_{\Delta x_{\chi}})\, A^\nu(x,N_{\Delta
x_{\chi}})=\nonumber \\=\left(\mathbf{\frac{\Delta}{\Delta_
{N_{\Delta x_\alpha}}}} \, \delta^\mu_\nu + \Gamma^\mu_{\nu\alpha}(x,N_{\Delta x_{\chi}})\right) \, A^\nu(x,N_{\Delta x_{\chi}}), \nonumber \\
\widetilde{A}_{^\mu;\alpha} \equiv \widetilde{D}_\alpha
A_\mu(x,N_{\Delta x_{\chi}})= \mathbf{\frac{\Delta}{\Delta_
{N_{\Delta x_\alpha}}}} A_\mu(x,N_{\Delta x_{\chi}})+
\Gamma^\nu_{\mu\alpha}(x,N_{\Delta x_{\chi}})\, A_\nu(x,N_{\Delta
x_{\chi}})=\nonumber
\\=\left(\mathbf{\frac{\Delta}{\Delta_ {N_{\Delta x_\alpha}}}} \,
\delta^\nu_\mu + \Gamma^\nu_{\mu\alpha}(x,N_{\Delta
x_{\chi}})\right) \, A_\nu(x,N_{\Delta x_{\chi}}),
\end{eqnarray}
where the operator $\mathbf{\Delta/\Delta_ {N_{\Delta x_\alpha}}}$
is taken from the formula (\ref{Div1.4new})in which
$N_{x_\mu}=N_{\Delta x_\mu}$, in accordance with {\bf Note 2.1}
from  Section 2.
\\
\\Similarly, for the {\it Riemann tensor} in a continuous theory we have
\cite{Einst1},\cite{Akhm}:
\begin{eqnarray}\label{Grav1.5.5}
{R^{\mu}}_{\nu\alpha\beta}(x) \equiv \partial_\alpha
\Gamma^\mu_{\nu\beta}(x)-\partial_\beta \Gamma^\mu_{\nu\alpha}(x)
+\Gamma^\mu_{\gamma\alpha}(x) \, \Gamma^\gamma_{\nu\beta}(x) -
\Gamma^\mu_{\gamma\beta}(x)\, \Gamma^\gamma_{\nu\alpha}(x).
\end{eqnarray}
With the use of formula (\ref{Grav1.5.3M}), we can get the
corresponding {\bf measurable} analog, i.e. the quantity
${R^{\mu}}_{\nu\alpha\beta}(x,N_{x_{\chi}})$ \cite{Shalyt-ASTP6}.
\\In a similar way we can obtain the {\bf measurable} variant of
{\it Ricci tensor}, $R_{\mu\nu}(x,N_{x_{\chi}}) \equiv
{R^\alpha}_{\mu\alpha\nu}(x,N_{x_{\chi}})$ , and the {\bf
measurable} variant of {\it Ricci scalar}: \\$R(x,N_{x_{\chi}})
\equiv R_{\mu\nu}(x,N_{x_{\chi}})\, g^{\mu\nu}(x,N_{x_{\chi}})$
\cite{Shalyt-ASTP6}.
\\So, for the {\it Einstein Equations} ($\mathcal{EE}$) in a continuous theory
\cite{Einst1},\cite{Akhm}
\begin{eqnarray}\label{Eeq}
 R_{\mu\nu} - \frac12 \, R \, g_{\mu\nu} - \frac12
\, \Lambda \, g_{\mu\nu} = 8\, \pi \, G \, T_{\mu\nu}
\end{eqnarray}
we can derive their {\bf measurable} analog, for short denoted as
($\mathcal{EEM}$) or {\it Einstein Equations Measurable}
\cite{Shalyt-ASTP6}:
\begin{eqnarray}\label{Eeq-M}
 R_{\mu\nu}(x,N_{x_{\chi}}) - \frac12 \,R(x,N_{x_{\chi}})\, g_{\mu\nu}(x,N_{x_{\chi}}) - \frac12
\, \Lambda(x,N_{x_{\chi}}) \, g_{\mu\nu}(x,N_{x_{\chi}}) \approx
\nonumber \\ \approx 8\, \pi \, G\, T_{\mu\nu}(x,N_{x_{\chi}}),
\end{eqnarray}
where $G$ -- Newton’s gravitational constant.
\\For correspondence with a continuous theory,
the following passage to the limit must take place for all the
points $x$:
\begin{eqnarray}\label{Cosm}
\lim\limits_{|N_{x_{\chi}}|\rightarrow \infty}
\Lambda(x,N_{x_{\chi}})= \Lambda,
\end{eqnarray}
where the cosmological constant  $\Lambda$ is taken from formula
(\ref{Eeq}).
\\Moreover, for high $|N_{x_{\chi}}|$, the quantity  $\Lambda(x,N_{x_{\chi}})$
should be practically independent of the point $x$, and we have
\begin{eqnarray}\label{Cosm2}
\Lambda(x,N_{x_{\chi}})\approx \Lambda(x^{'},N^{'}_{x^{'}_{\chi}})
\approx\Lambda,
\end{eqnarray}
where $x\neq x^{'}$ and  $|N_{x_{\chi}}|\gg
1,|N^{'}_{x^{'}_{\chi}}|\gg 1. $
\\Actually, it is clear that formula (\ref{Cosm}) reflects the fact that ($\mathcal{EEM}$)
given by formula (\ref{Eeq-M}) represents {\bf deformation} of the
Einstein equations ($\mathcal{EE}$) (\ref{Eeq}) in the sense of
the Definition given in \cite{Fadd} with the deformation parameter
$N_{x_{\chi}}$ (or $1/N_{x_{\chi}}$), and we have
\begin{eqnarray}\label{Deform1}
\lim\limits_{|N_{x_{\chi}}|\rightarrow \infty} \mathcal{EEM}= \mathcal{EE}\nonumber \\
{\rm or} \quad {\rm same} \quad
\lim\limits_{1/|N_{x_{\chi}}|\rightarrow 0}
\mathcal{EEM}=\mathcal{EE}.
\end{eqnarray}
We denote this deformation as $\mathcal{EEM}[N_{x_{\chi}}]$. Since
at low energies $E\ll E_P$ and to within the known constants we
have $\ell/N_{x_{\chi}}=p_{N_{x_{\chi}}}$, the following
deformations of $\mathcal{EE}$ are equivalent to
\begin{eqnarray}\label{Deform2}
\mathcal{EEM}[N_{x_{\chi}}]\equiv \mathcal{
EEM}[p_{N_{x_{\chi}}}].
\end{eqnarray}
So, on passage from $\mathcal{EE}$ to the {\bf measurable}
deformation of $\mathcal{EEM}[N_{x_{\chi}}]$ (or identically
$\mathcal{EEM}[p_{N_{x_{\chi}}}]$) we can find solutions for the
gravitational equations on the metric bundle
$\emph{B}_{g,N_{x_{\chi}}}\doteq g_{\mu\nu}(x,\{N_{x_{\chi}}\})$
(formula (\ref{GravM.1Bund})).

\subsection{Measurability and Least Action Principle  in Gravity}

{\bf 5.2.1.} First, we consider the variational principle in a
unidimensional case in the {\bf measurable} form. Next, we use the
terminology and notation of (\cite{Dubr}, paragraph 31).
\\ Let $S[\gamma]=\int_P^QL(\textbf{x},\dot{\textbf{x}},t)dt;\textbf{x}=\{x^i\},i=1,...,n$
be the action for the smooth curves $\gamma$ specified as  $\gamma:
x^i=x^i(t), a\leq t\leq b$, where the ends $a$ and $b$ are fixed and $\gamma$ connect the fixed pair of points
$P=(x^i_1)$,$Q=(x^i_2)$.
\\ Then the transition from a continuous consideration to the {\bf measurable} one is realized as
\begin{eqnarray}\label{Act-1.1}
S[\gamma]=\int_P^QL(\textbf{x},\dot{\textbf{x}},t)dt\rightarrow
S_{\bf
1/N_{t}}[\gamma]=\sum_P^QL(\textbf{x}(t),\mathbf{\frac{\Delta}{\Delta_
{N_t}}}\textbf{x}(t),t)\frac{1}{N_t}.
\end{eqnarray}
Here $\textbf{x},\dot{\textbf{x}}$ are considered as ''{\bf
measurable}'' functions of $t$ (i.e. their infinitesimal
variations at the point $t$ are of the form $1/N_t$), where $N_t$
-  high (in its absolute value)integer, or $|N_t|\gg 1$.
\\ It is clear that for high $|N_t|$ we have
\begin{eqnarray}\label{Act-1.1*}
S[\gamma]=\int_P^QL(\textbf{x},\dot{\textbf{x}},t)dt\approx S_{\bf
1/N_{t}}[\gamma]=\sum_P^QL(\textbf{x}(t),\mathbf{\frac{\Delta}{\Delta_
{N_t}}}\textbf{x}(t),t)\frac{1}{N_t}.
\end{eqnarray}
\\
\\ In the continuous case for great integer values of $N,|N|\gg 1$ we have
\begin{eqnarray}\label{Act-1.2}
\lim\limits_{\epsilon\approx
0}\frac{S[\gamma+\epsilon\eta]-S[\gamma]}{\epsilon}
\mapsto\lim\limits_{|N|\gg
1}\frac{S[\gamma+\frac{1}{N}\eta]-S[\gamma]}{1/N}
=\mathbf{\frac{\Delta}{\Delta_
{N}}}S[\gamma+\frac{1}{N}\eta]|_{|N|\gg 1},\nonumber \\
\lim\limits_{\epsilon\rightarrow
0}\frac{S[\gamma+\epsilon\eta]-S[\gamma]}{\epsilon}=\frac{d}{d\epsilon}S[\gamma+\epsilon\eta]|_{\epsilon=0}
=\lim\limits_{N\rightarrow
\infty}\frac{S[\gamma+\frac{1}{N}\eta]-S[\gamma]}{1/N}=\mathbf{\frac{\Delta}{\Delta_
{N}}}S[\gamma+\frac{1}{N}\eta]|_{N=\infty},\nonumber \\
\frac{d}{d\epsilon}S[\gamma+\epsilon\eta]|_{\epsilon=0}\approx
\mathbf{\frac{\Delta}{\Delta_
{N}}}S[\gamma+\frac{1}{N}\eta]|_{|N|\gg 1}.
\end{eqnarray}
Consequently, the equation
\begin{eqnarray}\label{Act-1.2.1}
\frac{d}{d\epsilon}S[\gamma+\epsilon\eta]|_{\epsilon=0}\equiv 0
\end{eqnarray}
leads to
\begin{eqnarray}\label{Act-1.2.2}
\mathbf{\frac{\Delta}{\Delta_
{N}}}S[\gamma+\frac{1}{N}\eta]|_{|N|\gg 1}\approx 0
\end{eqnarray}
and hence the variational derivative $S[\gamma]$ in the continuous case
\begin{eqnarray}\label{Act-1.3.1Cont}
\frac{\delta S}{\delta x^i}=\frac{\partial L}{\partial
x^i}-\frac{d}{dt}\frac{\partial L}{\partial\dot{x}^i}
\end{eqnarray}
generates the  {\bf measurable} variational derivative
\begin{eqnarray}\label{Act-1.3.1.Meas}
\frac{\delta_{\bf \{\hat{N}\}} S}{\delta
x^i}\doteq\frac{\Delta}{\Delta_ {N_{x^i}}} L-\frac{\Delta}{\Delta_
{N_{t}}}(\frac{\Delta}{\Delta_ {N_{\dot{x}^i}}}L),
\end{eqnarray}
where ${\bf \{\hat{N}\}}$--set of integer numbers sufficiently high in absolute value
${\bf \{\hat{N}\}}\doteq (N,N_{t},N_{x^i},N_{\dot{x}^i})$.
\\
\\ Due to the fact that absolute values of all the integer
numbers $N,N_{t},N_{x^i},N_{\dot{x}^i}$ may be taken no matter how
high, it holds true that
$\frac{\Delta}{\Delta_{{\bf\{\hat{N}\}}}}\approx
\frac{\Delta}{\Delta_{{\bf\{-\hat{N}\}}}}$.
\\
\\ From the condition
\begin{eqnarray}\label{Act-1.3.2Cont}
\frac{\delta S}{\delta x^i}=0
\end{eqnarray}
it follows that
\begin{eqnarray}\label{Act-1.3.2Meas}
\frac{\delta_{\bf \{\hat{N}\}}S}{\delta x^i}\approx 0,
\end{eqnarray}
where, as demonstrated in Section 4, the sign $\approx$ denotes
however high accuracy for sufficiently high $|{\bf\{\hat{N}\}}|$.
\\It should be noted that the formulae (\ref{Act-1.3.1.Meas}),
(\ref{Act-1.3.2Meas}) for the {\bf measurable} variational
derivative are easily derived from proof of the corresponding
theorem in the continuous case   (theorem 1 of paragraph 31 in
\cite{Dubr})with substitution of the sum for the integral and with
replacement of all derivatives in the continuous consideration by
their {\bf measurable} analogs in the right-hand side of formulae
(\ref{Act-1.1}) (\ref{Act-1.1*}):
\begin{eqnarray}\label{Act-1.4}
\int_a^b\mapsto\sum_a^b;dt\mapsto\frac{t}{N_t};\frac{d}{dt}\mapsto\frac{\Delta}{\Delta_
{N_t}};\frac{\partial}{\partial x^i}\mapsto\frac{\Delta}{\Delta_
{N_x^i}},...
\end{eqnarray}
Note that, as all the absolute values of the integers
$N,N_t,N_{x^i},...$ used in a {\bf measurable} consideration are
sufficiently high, it is obvious that a {\bf measurable}  analog
of the ''integration by parts'' at a however high accuracy in this
consideration is the case in the right side of formula
(\ref{Act-1.1*}) when using the substitution indicated in formula
(\ref{Act-1.4})  in the process of proof. And in the final variant
we have the formula (\ref{Act-1.3.2Meas}).
\\
\\{\bf 5.2.2.} In principle, consideration of the multidimensional case is no different
from that of the unidimensional one. In this case in the
continuous pattern formula for the variational derivative
(\ref{Act-1.3.1Cont}) is replaced by the formula (\cite{Dubr},
paragraph 37):
\begin{eqnarray}\label{Act-1.3.1*Cont}
\frac{\delta I}{\delta f^i}=\frac{\partial L}{\partial
f^i}-\sum_{\alpha=1}^n\frac{\partial}{\partial
x^\alpha}(\frac{\partial L}{\partial f_{x^\alpha}^i}),(1\leq i\leq
k),
\end{eqnarray}
where\begin{eqnarray}\label{Act-1.3.1*1Cont}
I[f]=\int_D L(x^\beta;f^i,f_{x^\alpha}^i)d^n x.
\end{eqnarray}
Here $D$-$n$-dimensional region, $d^nx$--n-dimensional form of the
volume in $\mathcal{R}^n$ and $f_{x^\alpha}^i(x^\beta)\doteq
\frac{\partial}{\partial x^\alpha}(f^i(x^\beta)).$
\\
\\ Accordingly, in the multidimensional {\bf measurable}
variant formula (\ref{Act-1.3.1.Meas})is replaced by
\begin{eqnarray}\label{Act-1.3.1*Meas}
\frac{\delta_{{\bf \{\hat{N}\}}} I}{\delta f^i}=\frac{\Delta
L}{\Delta_{N_{f^i}}}-\sum_{\alpha=1}^n\frac{\Delta}{\Delta_{N_
{x^\alpha}}}(\frac{\Delta L}{\Delta_{N_{f_{x^\alpha}^i}}}),(1\leq
i\leq k),
\end{eqnarray}
where ${\bf \{\hat{N}\}}$--set of integer numbers sufficiently high in absolute value
${\bf \{\hat{N}\}}\doteq
(N,N_{x^\alpha},N_{f^i},N_{f_{x^\alpha}^i}).$
\\Then the condition $\delta I/\delta f^i=0$ in the {\bf measurable} format
is equivalent to the condition
\begin{eqnarray}\label{Var1-Meas}
\frac{\delta_{\bf\{\hat{N}\}}I}{\delta f^i}\approx 0,(1\leq i\leq
k).
\end{eqnarray}
The functional $I[f]$ (formula (\ref{Act-1.3.1*1Cont})) is associated with the sum
\begin{eqnarray}\label{Act-1.3.1*Mes}
I_{\bf 1/N_{x^\alpha}}[f]=\sum_D
L(x^\beta;f^i,f_{N_{x^\alpha}}^i)(\frac{\ell}{N_{x^\alpha}})^n,
\end{eqnarray}
where all the notation is taken from formula
(\ref{Act-1.3.1*1Cont}), with substitution of
$f_{N_{x^\alpha}}^i(x^\beta)=\frac{\Delta
f^i(x^\beta)}{\Delta_{N_{x^\alpha}}}$   for
$f_{x^\alpha}^i(x^\beta)=\frac{\partial}{\partial
x^\alpha}(f^i(x^\beta))$ and of the {\bf measurable} quantity
$(\ell/N_{x^\alpha})^n$ for the volume form $d^n x$. (This means
that in the standard expression for the form of the
$n$-dimensional volume $d^n x=dx^1\wedge...\wedge dx^n$
(\cite{Dubr}, paragraph 37) the quantity $dx^\alpha$ is replaced
by the quantity $\ell/N_{x^\alpha}$ as at the end of  Section 4 in
formulae (\ref{Trans-M}), (\ref{Meas5}), etc.).
\\ Then in a multidimensional case at sufficiently high integer
numbers $|N_{x^\alpha}|$ for formula (\ref{Act-1.1*}) the
following formula is its obvious analog:
\begin{eqnarray}\label{Act-1.1*Mes}
I[f]=\int_D L(x^\beta;f^i,f_{x^\alpha}^i)d^n x\approx I_{\bf
1/N_{x^\alpha}}[f]=\sum_D
L(x^\beta;f^i,f_{N_{x^\alpha}}^i)(\frac{\ell}{N_{x^\alpha}})^n.
\end{eqnarray}
Similar to the unidimensional case, a {\bf measurable} variant of
the ''integration by parts''  is valid for the right-hand side of
formula(\ref{Act-1.1*Mes}), with replacement of  all derivatives
by their {\bf measurable}  analogs as it has been performed in
formula (\ref{Act-1.3.1*Meas}) compared with
(\ref{Act-1.3.1*Cont}).
\\
\\{\bf 5.2.3.} It is known that the gravitational action $S_{EH}$
in a continuous consideration is of the form
\cite{Land1}--\cite{Akhm}
\begin{eqnarray}\label{EHaction}
 S_{EH} = - \frac{1}{16 \pi G} \, \int
d^4x \, \sqrt{|g|} \, \left(R + \Lambda\right)+
S_M\left(g_{\mu\nu},{\it matt}\right).
\end{eqnarray}
Then, proceeding from the above-mentioned results,
in a {\bf measurable} variant it is associated with the quantity
\begin{eqnarray}\label{EHaction-M}
 S_{EH}(N_{x_{\chi}}) = - \frac{1}{16 \pi G} \, \sum
\Delta_{(N_{x_{\chi}})}\Omega \, \sqrt{|g(N_{\Delta x^{\chi}})|}
\, \left(R(x,N_{x_{\chi}}) + \Lambda(x,N_{x_{\chi}})\right)+
\nonumber \\
+S_M\left(g_{\mu\nu}(x,N_{x_{\chi}}),{\it matt}\right),
\end{eqnarray}
where the volume element in a {\bf measurable} variant
$\Delta_{(N_{x_{\chi}})}\Omega$ is taken from formulae
(\ref{Div1.6})--(\ref{Trans-M}),  the quantity $g(N_{\Delta
x^{\chi}})$ -- from the definition and formula (\ref{Meas3.2M3})
in Section 4 ; $R(x,N_{x_{\chi}})$ and $\Lambda(x,N_{x_{\chi}})$
are taken from formulae in Subsection 5.1. Finally,
$S_M\left(g_{\mu\nu}(x,N_{x_{\chi}}),{\it matt}\right)$ -- term in
the action corresponding to the {\bf measurable} form of matter
fields.
\\In canonical gravity the least action
principle is associated with formula \cite{Land1}--\cite{Akhm}
\begin{eqnarray}\label{minEH}
0 = \delta_g S_{EH} = -\frac{1}{16\, \pi \, G} \, \delta_g \, \int
d^4x \, \sqrt{|g|}\, \left(g^{\mu\nu} \, R_{\mu\nu} +
\Lambda\right) + \delta_g S_M = \nonumber \\ = - \frac{1}{16\,
\pi\, G} \, \int d^4x \, \left[\left(\delta \sqrt{|g|}\right) \,
\left(R+\Lambda\right) + \sqrt{|g|}\, \left(\delta
g^{\mu\nu}\right) \, R_{\mu\nu} + \sqrt{|g|} \, \left(\delta
R_{\mu\nu}\right) \, g^{\mu\nu}\right] + \delta_g S_M,
\end{eqnarray}
where $\delta_g S \equiv \left[S(g+\delta g) - S(g)\right]_{linear
\,\, in \,\, \delta g}$, and in the extremum of $S$ we have that
$\delta_g S = 0$.
\\ Due to the results from points {\bf 5.2.1.} and {\bf 5.2.2.}
in a {\bf measurable} consideration, formula (\ref{minEH}) is associated with the expression
\begin{eqnarray}\label{minEH-M}
0 \approx \delta_{g,{\bf N}} S_{EH}(N_{x_{\chi}}) = -\frac{1}{16\,
\pi \, G} \, \delta_{g,{\bf N}}\, \sum
\Delta_{(N_{x_{\chi}})}\Omega \, \sqrt{|g(N_{\Delta
x^{\chi}})|}(g^{\mu\nu}(x,N_{x_{\chi}}) \,
R_{\mu\nu}(x,N_{x_{\chi}})+
 \nonumber \\+\Lambda(x,N_{x_{\chi}})) +\nonumber \\
+\delta_{g,{\bf N}} S_M = \nonumber \\ = - \frac{1}{16\, \pi\, G}
\, \sum \Delta_{(N_{x_{\chi}})}\Omega \,[(\delta_{g,{\bf N}}
\sqrt{|g(N_{\Delta x^{\chi}})|}) \, (R(x,N_{x_{\chi}})+\Lambda(x,N_{x_{\chi}})) + \nonumber \\
+ \sqrt{|g(N_{\Delta x^{\chi}})|}\, (\delta_{g,{\bf N}}
g^{\mu\nu}(x,N_{x_{\chi}})) \, R_{\mu\nu}(x,N_{x_{\chi}}) +\nonumber \\
+\sqrt{|g(N_{\Delta x^{\chi}})|} \, (\delta_{g,{\bf N}}
R_{\mu\nu})(x,N_{x_{\chi}}) \, g^{\mu\nu}(x,N_{x_{\chi}})] +
\delta_{g,{\bf N}} S_M,
\end{eqnarray}
where, respectively,$\delta_{g,{\bf N}}\equiv [S(g+\frac{1}{{\bf
N}}\vartheta) - S(g)]_{linear \,\, in \,\,\frac{1}{{\bf
N}}\vartheta}$, the number ${\bf N}$ -- integer, $|{\bf N}|\gg 1$,
and in the extremum of $S$ we have that $\delta_{g,{\bf N}} S
\approx 0$ as in formulae (\ref{Act-1.2}),(\ref{Act-1.2.2}). In
this case, since the energies under consideration are low,
according to formulae (\ref{Cosm})and (\ref{Cosm2}),
$\delta_{g,{\bf N}}\Lambda(x,N_{x_{\chi}})\approx 0$ is satisfied.
\\The well-known formula for $\delta_g\sqrt{|g|}$ in the continuous pattern of a theory
\cite{Land1},\cite{Akhm}
\begin{eqnarray}\label{detg}
\delta_g\sqrt{|g|} = - \frac12 \sqrt{|g|} \, g_{\mu\nu} \,
\delta_g g^{\mu\nu}
\end{eqnarray}
in a {\bf measurable} variant for integers sufficiently high in
absolute value ${\bf N}$ corresponds to the formula
\begin{eqnarray}\label{detg-M}
\delta_{g,{\bf N}}\sqrt{|g(N_{\Delta x^{\chi}})|} \approx -
\frac12 \sqrt{|g(N_{\Delta x^{\chi}})|} \,
g_{\mu\nu}(x,N_{x_{\chi}}) \, \delta_{g,{\bf N}}
g^{\mu\nu}(x,N_{x_{\chi}})
\end{eqnarray}
to no matter how high an accuracy. In \cite{Shalyt-ASTP7} it has
been shown that the {\it strong principle of equivalence} is valid
for gravity in the {\bf measurable} form. This can be demonstrated
easily as within a random small neighborhood of any point there is
even smaller neighborhood specified in terms of {\bf measurable}
quantities.
\\In the continuous pattern from this property
it follows that in a sufficiently small neighborhood of the
arbitrary point $x_0$ we have
\begin{eqnarray}\label{LMRS}
 g_{\mu\nu}(x_0) = \eta_{\mu\nu}, \quad {\rm and}
\quad \Gamma^\alpha_{\beta\gamma} (x_0) = 0.
\end{eqnarray}
In the {\bf measurable} form the relation of (\ref{LMRS}) is
associated with the formula
\begin{eqnarray}\label{LMRS-M}
 g_{\mu\nu}(x_0,N_{(x_0){\chi}}) = \eta_{\mu\nu,}(x_0,N_{(x_0){\chi}}), \quad {\rm and}
\quad \Gamma^\alpha_{\beta\gamma}(x_0,N_{(x_0){\chi}})\approx 0.
\end{eqnarray}
Initially, the first relation in formula (\ref{LMRS-M}) should have been written as
$ g_{\mu\nu}(x_0,N_{(x_0){\chi}}) =
\eta_{\mu\nu,}(x_0,N'_{(x_0){\chi}})$, i.e.
sets of the integers high in absolute value $N_{(x_0){\chi}}$ and $N'_{(x_0){\chi}}$ should be different.
But we can demonstrate (see Section 3 in \cite{Shalyt-ASTP7}) that,
for sufficiently high absolute values of the integers from the sets
$N_{(x_0){\chi}}$, they may be considered as identical.
\\Without loss of generality, we can assume that the initial set of numbers
$N_{(x_0){\chi}}$ is exactly so.
\\In this way, using formula (\ref{LMRS-M}) and replacing the required formula
in the continuous pattern by the corresponding analogs in the {\bf
measurable} form, we obtain the  {\it Least Action Principle for
Gravity} in a {\bf measurable} variant.
\\ Specifically, in the continuous pattern
we have the equation (for example, formula (70) in \cite{Akhm})
\begin{eqnarray}\label{Bound-1}
\int_{\cal M} d^4x \, \sqrt{|g|} \, g^{\mu\nu} \, \delta
R_{\mu\nu} = \int_{\cal M} d^4x\, \sqrt{|g|} \, D_\mu \delta U^\mu
= \oint_{\partial {\cal M}} d\Sigma_\mu \, \delta U^\mu,
\end{eqnarray}
where ${\cal M}$ is the space--time manifold under consideration
and $\partial {\cal M}$ is its boundary, $d\Sigma_\mu$ is the
four--vector normal to $\partial {\cal M}$, whose modulus is the
infinitesimal volume element of $\partial {\cal M}$: $d\Sigma_\mu
= n_\mu \, \sqrt{g^{(3)}} \, d^3\xi$, where $n_\mu$ is the normal
vector to the boundary. And $g^{(3)} = \left|\det g_{ij}\right|$
is the determinant of the induced three--dimensional metric,
$g_{ij}, \,\, i = 1,2,3$, on the boundary $\partial {\cal M}$ and
$\xi$ are the corresponding coordinates parametrizing the
boundary, and $\delta U^\mu\doteq g^{\alpha\beta} \, \delta
\Gamma^\mu_{\alpha\beta} - g^{\alpha\mu} \, \delta
\Gamma^\beta_{\alpha\beta}.$
\\ In a {\bf measurable} consideration in formula (\ref{Bound-1}) the substitution takes place
\begin{eqnarray}\label{Bound-1M}
\int_{\cal M}\mapsto\sum_{\cal M},
d^4x\mapsto\Delta_{(N_{x_{\chi}})}\Omega, g^{\mu\nu}\mapsto
g_{\mu\nu}(x,N_{x_{\chi}}),g\mapsto g(N_{\Delta x^{\chi}}),\nonumber \\
\delta R_{\mu\nu}\mapsto \delta_{g,{\bf N}}
R_{\mu\nu}(x,N_{x_{\chi}}),\delta U^\mu\mapsto\delta_{\bf
N}U^\mu(x,N_{x_{\chi}})\doteq \nonumber \\
\doteq g^{\alpha\beta}(x,N_{x_{\chi}})\delta_{g,{\bf N}}
\Gamma^\mu_{\alpha\beta}(x,N_{x_{\chi}}) -
g^{\alpha\mu}(x,N_{x_{\chi}})\delta_{g,{\bf N}}
\Gamma^\beta_{\alpha\beta}(x,N_{x_{\chi}}),\nonumber \\
D_\mu\mapsto\widetilde{D}_\mu, \oint_{\partial {\cal
M}}\mapsto\sum_{\partial {\cal M}},d\Sigma_\mu\mapsto
\Delta\Sigma_\mu\doteq n_\mu \, \sqrt{g^{(3)}(x,N_{x_{i}}})
\,\prod\frac{\ell}{N_{x_{i}}}.
\end{eqnarray}
In the last line of formula (\ref{Bound-1M}) $\widetilde{D}_\mu$
taken from  (\ref{Grav1.6.Meas}),$\sum_{\partial {\cal M}}$ is
understood as a sum over the boundary $\partial {\cal M}$,and,
finally, $\Delta\Sigma_\mu$  is derived from $d\Sigma_\mu$ with
the already known substitution (in this case for the
three-dimensional case) $g^{(3)}\mapsto g^{(3)}(x,N_{x_{i}})$ and
$dx_i\mapsto\ell/N_{x_{i}}.$
\\ In the continuous pattern, due to the Stokes formula,
$\oint_{\partial {\cal M}} d\Sigma_\mu \, \delta U^\mu=0$. But as
$\sum_{\partial {\cal M}}\Delta\Sigma_\mu\delta_{\bf
N}U^\mu(x,N_{x_{\chi}})\approx \oint_{\partial {\cal M}}
d\Sigma_\mu \, \delta U^\mu$, in the {\bf measurable} form we have
$\sum_{\partial {\cal M}}\Delta\Sigma_\mu\delta_{\bf
N}U^\mu(x,N_{x_{\chi}})\approx 0$.
\\ From whence we obtain, in accordance with a continuous consideration,
the formula for a {\bf measurable} variant as follows:
\begin{eqnarray}\label{minEH-Mnew}
\delta_{g,{\bf N}} S_M - \frac{1}{16\, \pi\, G} \, \sum
\Delta_{(N_{x_{\chi}})}\Omega\sqrt{|g(N_{\Delta x^{\chi}})|}
[R_{\mu\nu}(x,N_{x_{\chi}})-\frac{1}{2}g_{\mu\nu}(x,N_{x_{\chi}})R(x,N_{x_{\chi}})
-\nonumber
\\-\frac{1}{2}g_{\mu\nu}(x,N_{x_{\chi}})\Lambda(x,N_{x_{\chi}})]\delta_{g,{\bf
N}} g^{\mu\nu}(x,N_{x_{\chi}}) \approx 0.
\end{eqnarray}
In canonical gravity the variation $\delta_g S_M$ takes the form
\cite{Einst1},\cite{Akhm}
\begin{eqnarray}\label{delSM}
 \delta_g S_M = \delta \int d^4x \, \sqrt{|g|} \,
{\cal L} = \int d^4x \, \left[\frac{\partial {\cal L}}{\partial
g^{\mu\nu}} \, \delta g^{\mu\nu} \, \sqrt{|g|} + {\cal L} \,
\delta \sqrt{|g|}\right] = \nonumber \\ = \int d^4x \, \sqrt{|g|}
\, \left[\frac{\partial {\cal L}}{\partial g^{\mu\nu}} - \frac12
\, {\cal L} \, g_{\mu\nu}\right] \, \delta g^{\mu\nu} \equiv
\frac12 \, \int d^4x \, \sqrt{|g|} \, T_{\mu\nu} \, \delta
g^{\mu\nu},
\end{eqnarray}
where ${\cal L}$--Lagrangian for matter fields and the
corresponding energy-momentum tensor $T_{\mu\nu}$ is defined as
\begin{eqnarray}\label{delSM-T}
 T_{\mu\nu} = 2 \,\frac{\partial {\cal L}}{\partial g^{\mu\nu}} - {\cal L} \,
g_{\mu\nu}, \quad T_{\mu\nu} = T_{\nu\mu}.
\end{eqnarray}
Using the foregoing results, it is easy to derive a {\bf
measurable} variant  of (\ref{delSM}),i.e. $\delta_{g,{\bf N}}
S_M$, replacing all the required components in formula
(\ref{delSM}) by their {\bf measurable} analogs from formula
(\ref{Bound-1M}) in the assumption that ${\cal L}={\cal
L}(g^{\mu\nu}(x,N_{x^{\chi}}))$,$\frac{\partial {\cal L}}{\partial
g^{\mu\nu}}\mapsto\frac{\Delta {\cal
L}(g^{\mu\nu}(x,N_{x^{\chi}}))}{\Delta_{N}}$,where
\begin{eqnarray}\label{delSM-Tvar}
\frac{\Delta {\cal L}(g^{\mu\nu}(x,N_{x^{\chi}}))}{\Delta_{N}}
\doteq \frac{{\cal
L}(g^{\mu\nu}(x,N_{x^{\chi}})+\frac{1}{N}g^{\mu\nu}(x,N_{x^{\chi}}))-{\cal
L}(g^{\mu\nu}(x,N_{x^{\chi}}))}
{\frac{1}{N}g^{\mu\nu}(x,N_{x^{\chi}})}.
\end{eqnarray}
It is clear that in this case formula (\ref{delSM-T}) is replaced by
\begin{eqnarray}\label{delSM-TM}
 T_{\mu\nu,\bf
N}\approx 2 \,\frac{\Delta {\cal
L}(g^{\mu\nu}(x,N_{x^{\chi}}))}{\Delta_{N}} - {\cal L} \,
g_{\mu\nu}(x,N_{x_{\chi}}),\nonumber \\ T_{\mu\nu,\bf N} \approx
T_{\nu\mu,\bf N}.
\end{eqnarray}
In a continuous consideration we have the identity
\begin{eqnarray}\label{Covar}
D^\mu T_{\mu\nu} = 0
\end{eqnarray}
that in a{\bf measurable} variant  takes the form
\begin{eqnarray}\label{Covar-M}
\widetilde{D}^\mu T_{\mu\nu,\bf N}\approx 0,
\end{eqnarray}
$\widetilde{D}^\mu $ -- covariant derivative in a {\bf
measurable} form obtained from  formula (\ref{Grav1.6.Meas}).
\\The identity (\ref{Covar}) may be derived from
the equation (for example, see formula (79) in \cite{Akhm})
\begin{eqnarray}\label{Var-new1}
0 \equiv \delta_{\epsilon} S_M = \int_{\cal M} d^4 x\, \sqrt{|g|}
\, T_{\mu\nu} \, D^\mu \epsilon^\nu = \nonumber \\ = \int_{\cal M}
d^4x\, \sqrt{|g|} \, D^\mu\left(T_{\mu\nu} \, \epsilon^\nu\right)
- \int_{\cal M} d^4x \, \sqrt{|g|} \, \epsilon^\nu \left(D^\mu \,
T_{\mu\nu}\right) = \nonumber \\ = \oint_{\partial {\cal M}}
d\Sigma^\mu \, T_{\mu\nu} \, \epsilon^\nu - \int_{\cal M} d^4x \,
\sqrt{|g|}\, \epsilon^\nu \, \left(D^\mu \, T_{\mu\nu}\right),
\end{eqnarray}
where  $T_{\mu\nu} = T_{\nu\mu}$, on going from the first to the
second line, the integration by parts is performed, and then for
the first term in the second line the Stokes theorem is used. It
is assumed that $\epsilon^\mu(x)$ is a small vector field such
that $\bar{x}^\mu = x^\mu + \epsilon^\mu(x)$,
$\epsilon^\mu(x)|_{\partial {\cal M}}=0$ for all $\mu$ , and we
have
\begin{eqnarray}\label{g1}
g^{\mu\nu}\left(\bar{x}\right) = g^{\mu\nu}\left(x +
\epsilon\right) \approx g^{\mu\nu}(x) + \partial_\alpha
g^{\mu\nu}(x) \, \epsilon^\alpha.
\end{eqnarray}
Then, in the assumption that $\epsilon^\mu(x)$ is a small {\bf
measurable} vector field (i.e. a small vector field expressed in terms of
{\bf measurable} quantities), a {\bf
measurable} variant of formula (\ref{g1}) may be obtained as follows:
\begin{eqnarray}\label{g1-M}
g^{\mu\nu}(\bar{x},N_{\bar{x}^{\chi}})= g^{\mu\nu}(x
+\epsilon,N_{(x + \epsilon)^{\chi}}) \approx
g^{\mu\nu}(x,N_{x^{\chi}}) + \Delta_{\alpha}
g^{\mu\nu}(x,N_{x^{\chi}}) \, \epsilon^\alpha,
\end{eqnarray}
where $\Delta_{\alpha}$ is taken from formula (\ref{Grav1.5d2}) in
Section 3.
\\Next, by substitution of all the components in equation (\ref{Var-new1})
according to formula (\ref{Bound-1M}) and with replacement of $
T_{\mu\nu}\mapsto by T_{\mu\nu,\bf N}$, of the sign $=$ by
$\approx$ and assuming $\epsilon^\alpha$ to be a small {\bf
measurable} vector field, we can obtain a  {\bf measurable} copy
of the equation (\ref{Var-new1}).
\\In this case the Stokes theorem and the ''integration by parts'',
to a high accuracy, are fulfilled now not for the integrals but
for the sums corresponding to these integrals and close in the
values, as it has been already noted in points   {\bf 5.2.1.} and
{\bf 5.2.2.}.
\\The ''integration by parts'' is of the form
\begin{eqnarray}\label{Var-new2m}
\sum_{\cal M} \Delta_{(N_{x_{\chi}})}\Omega\, \sqrt{|g(N_{\Delta
x^{\chi}})|} \,T_{\mu\nu,\bf N} \,
\widetilde{D}^\mu \epsilon^\nu \approx \nonumber \\
\approx\sum_{\cal M} \Delta_{(N_{x_{\chi}})}\Omega\,
\sqrt{|g(N_{\Delta x^{\chi}})|} \,
\widetilde{D}^\mu\left(T_{\mu\nu,\bf N}\, \epsilon^\nu\right)-\nonumber \\
-\sum_{\cal M} \Delta_{(N_{x_{\chi}})}\Omega\, \sqrt{|g(N_{\Delta
x^{\chi}})|}\, \epsilon^\nu \left(\widetilde{D}^\mu \,
T_{\mu\nu,\bf N}\right) \approx 0.
\end{eqnarray}
Naturally assuming that at a small {\bf measurable} vector
field $\epsilon^\mu$ for all $\mu$ the condition
$\epsilon^\mu(x)|_{\partial {\cal M}}=0$  is fulfilled, we obtain the formula (\ref{Covar-M}).
\\ As ${\bf N}$ satisfies the condition $|{\bf N}|\gg 1$, from formula
(\ref{Covar-M}) it follows that $T_{\mu\nu,\bf
N}$  is practically independent of ${\bf N}$, i.e. we have
\begin{eqnarray}\label{delSM-TMnew}
 T_{\mu\nu,\bf
N}\approx T_{\mu\nu},|{\bf N}|\gg 1,
\end{eqnarray}
and we come to a {\bf measurable} copy of the equation (\ref{Eeq} given by formula (\ref{Eeq-M}).
\\
\\ So, the principal inference from this section is as follows:
\\{\it at low energies $E\ll E_p$ Gravity (GR) in the continuous space-time
may be redetermined in terms of the {\bf measurability} notion
using in  GR, instead of the limiting quantities, the finite
measurable quantities   which are close to them and determined by
the existing energies.
\\ Let us call this theory redetermined in terms of {\bf measurable} quantities the Measurable Gravity (or GRM).
GRM is a discrete theory close to GR but not identical to it.
Possible distinctions of GRM from GR are considered in the
subsequent section.
\\However, in Section 7 it is shown that in the proposed approach the {\bf measurable}
form of Einstein Equations may be correctly defined at high
energies $E\approx E_p$ as well. It should be noted that the
indicated form completely satisfies the Correspondence Principle,
i.e. in the limiting transition to low energies we involve
$\mathcal{EEM}$ from formula (\ref{Eeq-M}).}

\section{Physical Meaning of Measurability and Some of Its Possible Inferences}

{\bf 6.1.} The principal significance of introducing the {\bf
measurability} notion is in the fact that at low energies $E\ll
E_p$, instead of abstract, dependent on nothing small $\delta
x_\mu$ and infinitesimal $d x_\mu$ space-time increments, we
introduce the variations of $\ell/N_{x_\mu};|N_{x_\mu}|\gg
1$,where $N_{x_\mu}$-integer.
\\These variations are determined by the {\bf primarily measurable} momenta $p_{N_{x_\mu}}$,
i.e. by the particular energies. By virtue of their definition,
the {\bf primarily measurable} momenta $p_{N_{x_\mu}}$ , to a high
accuracy, determine the whole totality of the momenta (and their
variations) $P_{LE}$ at low energies.
\\ The ''construction material'' of a theory is changed considerably
from the abstract $\delta x_\mu,d x_\mu,dp_{x_i},dE$ in a
continuous consideration to $\ell/N_{x_\mu},p_{N_{x_\mu}}$ having
specific values in the {\bf measurable} pattern, as is directly
indicated in formulae  of Section 2  and in  {\bf Remark 1} at the
very ending of this section.
\\ The associated model is discrete but in this case, due to
formulae (\ref{Div1.4}),(\ref{Div1.4new}), it is very close to the
initial continuous theory. Here the words ''very close '' do not
mean that it is identical.
\\ Let us return to formulae (\ref{Eeq-M}) and (\ref{Deform1}) of Section 5.
As shown by these formulae, for high $|N_{x_\mu}|$
  {\it Einstein Equations Measurable} ($\mathcal{EEM}$) and
 {\it Einstein Equations} ($\mathcal{EE}$)
  are very close but in fact the solutions of these equations are differing.
 \\ Indeed, one metric $g_{\mu\nu}(x)$ forming a solution of $\mathcal{EE}$
and given by formula (\ref{Grav1.4}) is associated with the whole
class of metrics  that is a layer over  the ''point''
$g_{\mu\nu}(x)$ in the bundle (\ref{GravM.1Bund}) and is given by
formula (\ref{Grav1.5}). Then it is quite probable that some
points based on the bundle (\ref{GravM.1Bund}) at the particular,
quite natural conditions imposed on the numbers $N_{x_\mu}$ have
no prototype. Specifically, this primarily refers to the {\it
Closed Time-like Curves} (CTC) \cite{Godel}--\cite{Lobo} from
$\mathcal{EE}$, which clearly {\it are not the solutions} having
the physical meaning.
\\It should be recalled that the curve $\gamma$ is referred
to as the {\it time-like} curve if the norm of its tangent vector
$T^{\mu}=dx_\mu(t')/dt'$ is everywhere negative, i.e.
$g_{\alpha\beta}T^{\alpha}T^{\beta}<0.$ \cite{Einst1}. Here
$t'$--parameter along the curve.
\\ It is obvious that in a {\bf measurable} consideration we have the substitution
\begin{eqnarray}\label{Phys.1}
g_{\alpha\beta}(x)\mapsto
g_{\alpha\beta}(x,N_{x_{\chi}});T^{\mu}=\frac{dx_\mu(t')}{dt'}\mapsto
T_{N'_{t'}}^{\mu}=\frac{\Delta x_\mu(t')}{\Delta_{N'_{t'}}},
\end{eqnarray}
where $N'_{t'}$--integer that is high in absolute value.
\\In this way the formulation of the CTC problem in the {\bf measurable} case
is varied and begins to be dependent on  the parameters
$\{N_{x_{\chi}}\}$ (and also on the additional small discrete
parameter $1/N'_{t'}$ along the curve $\gamma$).
\\Clearly, other quantities, e.g., lengths of the curves,
in a {\bf measurable} variant are varied too. In particular, for
the {\it space-like} curve  the transition from the quantity of
length in a continuous variant $\mathcal{L}$ \cite{Einst1} to that
in a {\bf measurable} variant
$\mathcal{L}_{meas}(\{N_{x_{\chi}}\},N'_{t'})$ may be given by the
formula
\begin{eqnarray}\label{Comp-Curv}
\mathcal{L}=\int[g_{\mu\nu}T^{\mu}T^{\nu}]^{1/2}dt\mapsto\mathcal{L}_{meas}(N_{x_{\chi}},N'_{t'})=\nonumber \\
=\sum
[g_{\mu\nu}(x,N_{x_{\chi}})T_{N'_{t'}}^{\mu}T_{N'_{t'}}^{\nu}]^{1/2}\frac{1}{N'_{t'}}.
\end{eqnarray}
The formula similar to (\ref{Comp-Curv}) is written for the {\it
Time-like} curves with changing in (\ref{Comp-Curv}) of the sign
before $g_{\mu\nu}$ (and hence before
$g_{\mu\nu}(x,\{N_{x_{\chi}}\}$)from  ''+'' to ''-'' and with
replacement of $\mathcal{L}$ by the intrinsic time $\mathcal{T}$
(and hence replacement of $\mathcal{L}_{meas}$ by
$\mathcal{T}_{meas}$).The absence of CTC in the measurable variant
of gravity  means that, under certain conditions for collection
$\{N_{x_{\chi}}\}$ for  metrics $\tilde{g}_{\mu\nu}(x)$ in General
Relativity generating the Closed Time-like Curves we have no {\bf
prototype}  in the mapping (\ref{GravM.1Bund}).
\\ So, the introduction of {\bf measurability} allows for
''additional degrees of freedom '' in solution of the specific
problems of General Relativity, CTC problem in particular.
\\ At low energies $E\ll E_p$
all {\bf measurable} quantities are little different from the
corresponding quantities in a continuous theory. Specifically,
$\mathcal{L}\approx\mathcal{L}_{meas}$. Besides, it is obvious
that all the above mentioned  $N_{x_{\mu}}$ should have the upper
limit for their absolute values. As $\ell\propto l_p\approx
10^{-33}cm$, all $|N_{x_{i}}|;i=1,2,3$ have the {\it absolute}
upper limit of $\mathcal{N}\approx 10^{61}$:
\begin{eqnarray}\label{MOD-1}
1\ll|N_{x_{i}}|\leq\mathcal{N}.
\end{eqnarray}
The relation (\ref{MOD-1}) is valid because all $|N_{x_{i}}|$ are
determined by the {\it real} {\bf primarily measurable} space
quantities $N_{x_{i}}\ell$, but we understand that
$|N_{x_{i}}|\ell$  never exceeds a radius of the visible part of
the Universe $\mathcal{R}_{Univ}\approx
10^{28}cm\approx\mathcal{N}\ell.$ Similarly, the absolute {\it
upper} limit $\mathcal{N}_t$  may be introduced for $|N_{x_{0}}|$
as well.
\\ In reality the upper limits for $|N_{x_{i}}|;i=1,2,3$ and
$|N_{x_{0}}|$ are  considerably lower than $\mathcal{N}$  and
$\mathcal{N}_t$, respectively, and are dependent on the problem at
hand. For example,   as the atomic radius has the characteristic
scale $\approx 10^{-8}cm$, we can demonstrate that in this case
(\ref{MOD-1}) is changed by (\ref{MOD-2}):
\begin{eqnarray}\label{MOD-2}
1\ll|N_{x_{i}}|\leq\mathcal{N}_a\approx 10^{25}.
\end{eqnarray}
In the case of a nucleus, the radius of which is on the order of
$\approx 10^{-12}\div10^{-12}cm$ (\ref{MOD-2}), is changed by
(\ref{MOD-3})
\begin{eqnarray}\label{MOD-3}
1\ll|N_{x_{i}}|\leq\mathcal{N}_{nucl}\approx 10^{21}.
\end{eqnarray}
The formulae similar to (\ref{MOD-2}) and (\ref{MOD-3}) can be
derived for $N_{x_{0}}$ too.
\\
\\{\bf 6.2.} As in the well-known works by S.Hawking \cite{Hawk-new1}
--\cite{Hawk-new3} all the results have been obtained within the
scope of the semiclassical  approximation, seeking for a solution
of the above-mentioned problem is of primary importance. More
precisely, we must find, {\it how to describe thermodynamics and
quantum mechanics using the “language” of the {\bf measurable}
variant of gravity and what is the difference (if any) from the
continuous treatment in this case}.
\\To have a deeper understanding of the problem, we should know about the transformations
of the notion of quantum information for the {\bf measurable}
variant of gravity and quantum theory at low  $E\ll E_P$ and at
high $E\approx E_P$ energies. Possibly, a new approach to the
solution of the Information Paradox Problem \cite{Hawk-new1} will
offer a better insight.
\\ Because of this, it is very important to understand,
which distinctive features has a semiclassical  approximation in
terms of {\bf measurable} quantities. As the semiclassical
approximation is consideration of the material quantum fields
against the background of the classical space-time, it is
important to study a dynamics of the system behavior in terms of
{\bf measurability} at low energies $E\ll E_p$, i.e. for
significantly different $N_{x_i}.$ Specifically, of great
importance is to get the answer to the following question: if we
know system’s behavior for the energy $E\ll E_p$ that is
associated with the numbers $N_{x_i}$, what is the behavior of the
same system at higher energies $E';E\ll E'\ll E_p$ associated in
the {\bf measurable} form with the sets $N'_{x_i}$ so that we have
\begin{eqnarray}\label{Semic} 1\ll|N'_{x_{i}}|\ll|N_{x_{i}}|.
\end{eqnarray}
\\
\\{\bf 6.3.} Finally, the proposed approach from the start is quantum
in character due to the fact that the fundamental length $\ell$ is
proportional to the Planck length $\ell\propto l_P$ and includes
the whole three fundamental constants, the Planck constant $\hbar$
as well. Besides, it is naturally dependent on the energy scale:
sets of the metrics $g_{\mu\nu}(x,\{N_{x_{\chi}}\})$  with the
lowest value $|N_{x_{\chi}}|$ correspond to higher energies as
they correspond to the momenta $\{p_{N_{x_{\chi}}}\}$ which are
higher in absolute value. This is the case for all the energies
$E.$
\\However, minimal measurable increments for the energies $E\approx E_P$
are not of the form $\ell/N_{x_{\mu}}$ because the corresponding
momenta $\{p_{N_{x_{\chi}}}\}$ are no longer {\bf primary
measurable}, as indicated by the results in  Section 2.
\\So, in the proposed paradigm the problem of the ultraviolet generalization
of the low-energy {\bf measurable} gravity
$(\mathcal{EEM})[N_{x_{\chi}}]$ (formula (\ref{Eeq-M})) is
actually reduced to the problem: what becomes with the {\bf
primary measurable} momenta
$\{p_{N_{x_{\chi}}}\},|N_{x_{\chi}}|\gg 1$ at high Planck’s
energies.
\\Just this problem is studied in the following section.

\section{Transition to High Energies in the Framework of Measurable Gravity}

As shown in this work, within the concept of {\bf measurability},
at low energies $E\ll E_p$ all {\bf small} space-time variations
are determined by  the {\bf primarily measurable} momenta from
formula (\ref{Meas-D2.G}). However, a simple example from Section
2 points to the fact that at high energies $E\approx E_p$ this is
not the case. As we intend to construct a {\bf measurable} variant
of the theory at high energies $E\approx E_p$, in this case we
should obtain the following:
\\
\\A) some  discrete model for the integers $N_{x_\mu}$ so that
$|N_{x_\mu}|\approx 1$ (the condition ''$|N_{x_\mu}|\approx 1$''should
be included depending on the specific problem at hand);
\\
\\B) the above-mentioned discrete model, according to the {\bf Principle of Correspondence},
on going to low energies $E\ll E_p$ should lead to a {\bf
measurable} variant of gravity from Sections 3-5 to a high
accuracy.
\\
\\ Then it is natural to suppose that in such a discrete model all {\bf measurable} analogs
of the {\bf small} space-time variations should be determined by
some {\bf generalized measurable}momenta $p\doteq
p(N_{x_\mu});|N_{x_\mu}|\approx 1$ so that on passage to low
energies these momenta give{\bf primarily measurable} momenta from
formula (\ref{Meas-D2.G})
\\ What do they look like these {\bf generalized measurable}
momenta \\$p\doteq p(N_{x_\mu});|N_{x_\mu}|\approx 1$?
\\In a relatively simple case of GUP in Section 2 we have the answer.
And, using the fact that $(\mathcal{EEM})[N_{x_{\chi}}]\equiv
(\mathcal{EEM})[p_{N_{x_{\chi}}}]$ (\ref{Deform2}), based on the
results of Section 2, we can construct a correct high-energy
passage to the Planck energies $E\approx E_p$ \cite{Shalyt-ASTP6}
\begin{eqnarray}\label{Deform2H}
(\mathcal{EEM})[p_{N_{x_{\chi}}},|N_{x_{\chi}}|\gg 1]\mapsto
(\mathcal{EEM})[p_{N_{x_{\chi}}}(GUP),|N_{x_{\chi}}|\approx 1],
\end{eqnarray}
where $p_{N_{x_{\chi}}}(GUP)=\Delta p(\Delta x_{\chi},GUP)$
according to formula (\ref{root3.3})  of  Section 2. In this
specific case, we can construct the natural ultraviolet
generalization $(\mathcal{EEM})[p_{N_{x_{\chi}}},|N_{x_{\chi}}|\gg
1]\doteq(\mathcal{EEM})[p_{N_{x_{\chi}}}].$ The theoretical
calculations
\\$(\mathcal{EEM})[p_{N_{x_{\chi}}}(GUP),|N_{x_{\chi}}|\approx 1]$ derived
at Planck’s energies are obviously {\it discrete},{\bf
measurable}, and represent a high-energy deformation in the sense
of the \cite{Fadd} {\bf measurable} gravitational theory
$(\mathcal{EEM})[p_{N_{x_{\chi}}},|N_{x_{\chi}}|\gg 1].$
\\GUP in Section 2 (formula (\ref{U2})) is a partial case of a
''broader'' GUP resultant in the minimal length $\ell$
\cite{Kempf},\cite{Tawf}
\begin{equation}\label{Kempf-1}
\Delta x \Delta p \ge \hbar (1+\beta (\Delta p)^2 + \beta \langle
{\bf{p}} \rangle^2),
\end{equation}
where $\beta>0$.
\\Formula (\ref{Kempf-1}) gives rise to the absolutely
smallest uncertainty in the positions
\begin{equation}\label{Kempf-2}
\Delta x_0 = 2\hbar\sqrt{\beta}\doteq \ell.
\end{equation}
To attain conformity of the formulae (\ref{U2}) and
(\ref{Kempf-1}), in the right-hand side of (\ref{Kempf-1}) we use
before $\hbar$ the factor  1, instead of 1/2 in \cite{Kempf}.
\\ It is clear that (\ref{Kempf-1}) represents a more general form of GUP
than (\ref{U2}, at least due to the fact that the minimal length
$\ell$ involved with it not necessarily should be on the order of
the Planck length $l_p$, i.e. proportional to it with the factor
about 1. Nevertheless, we still assume in this work that the
minimal length  $\ell$ should be on the order of the Planck length
$\ell\propto l_p$ because all the previous results point to the
fact that for gravity the onset of a quantum regime begins at the
Planck energies. \cite{QG1}--\cite{Planck3.1}. But all the
calculation are valid for the random minimal length $\ell$ too.
\\ Assuming in (\ref{Kempf-1}) the equality $\langle
{\bf{p}}\rangle=0$ and $\Delta x$ {\bf primarily measurable}
quantity, (i.e. $\Delta x=N_{\Delta x}\ell$), we arrive at  a
formula for $\Delta p$ that is completely coincident with formula
(\ref{root3.3}) for the partial case of GUP (\ref{U2}).In this way
GUP, both in the partial and in the general cases, demonstrates
that   at high energies the {\bf primarily measurable} momenta are
inadequate for the construction of a theory and the {\bf
generalized measurable} momenta should be involved.
\\As noted in Section 2, the main target of the author is to construct
a theory at all energy scales  in terms  of {\bf generalized
measurable} (or same {\bf  measurable}) quantities.
\\In this theory the values of the physical quantity $\mathcal{G}$
may be represented by the numerical function $\mathcal{F}$ in the
following way:
\begin{eqnarray}\label{Meas-D*F.}
\mathcal{G}=\mathcal{F}(N_{i},N_{t},\ell)=\mathcal{F}(N_{i},N_{t},G,\hbar,c,\kappa),
\end{eqnarray}
where in the general case $N_{i},N_{t}$--integer numbers from the
formulae (\ref{Meas-D2.G}),(\ref{Meas-D3.}) and $G,\hbar,c$ are the
fundamental constants. The last equality in (\ref{Meas-D*F.}) is
determined by the fact that $\ell=\kappa l_p$ and
$l_p=\sqrt{G\hbar/c^{3}}$.
\\ As shown above, at low energies $E\ll E_p$ for the momenta we have $p_{N_{x_\mu}}=\mathcal{G}$
because the latter are {\bf primarily measurable} quantities,
formula (\ref{Meas-D*F.}) is simplified and may be derived with
the help of formula (\ref{Meas-D2.}).
\\ Let us assume that at high energies $E\approx E_p$ we have
a certain formula (\ref{Meas-D*F.}) for the {\bf generalized
measurable} momenta $p_{N_{x_\mu}}=\mathcal{G};|N_{x_\mu}|\approx
1$. As demonstrated above, this is the case, for example, for GUP
(\ref{U2})),(\ref{Kempf-1}). Then in a {\bf measurable} variant of
the theory, due to the  {\bf Principle of Correspondence}, these
{\bf generalized measurable} momenta at low energies $E\ll E_p$,
to a high accuracy, should lead to  the {\bf primarily measurable}
momenta
\begin{eqnarray}\label{Deform2Hnew2}
p_{N_{x_{\chi}}},(|N_{x_{\chi}}|\approx
1)\stackrel{|N_{x_{\chi}}|\approx 1 \rightarrow |N_{x_{\chi}}|\gg
1}{\Rightarrow}p_{N_{x_{\chi}}},(|N_{x_{\chi}}|\gg 1),
\end{eqnarray}
where momenta in the right-hand part of formula
(\ref{Deform2Hnew2}), i.e. $p_{N_{x_{\chi}}},(|N_{x_{\chi}}|\gg
1)$, are  the {\bf primarily measurable} momenta at low energies
$E\ll E_p$.
\\ By formulae (\ref{root3.3}) and (\ref{root3.2.}) of
Section 2 it has been shown that (\ref{Deform2Hnew2}) is valid for
GUP (\ref{U2})) and, proceeding from all the above, for a more
general form of GUP (\ref{Kempf-1}).
\\ Then, in accordance with formula (\ref{Deform2Hnew2}),
 in the general case the transition from high $E\approx E_p$ to low
energies for a {\bf measurable} variant  of gravity is given as the
low-energy deformation:
\begin{eqnarray}\label{Deform2Hnew}
\mathcal{(EEM)}[p_{N_{x_{\chi}}},|N_{x_{\chi}}|\approx 1]\mapsto
\mathcal{(EEM)}[p_{N_{x_{\chi}}},|N_{x_{\chi}}|\gg 1].
\end{eqnarray}
Next, in a quite natural way, we assume that in all the cases
for a {\bf measurable} variant of gravity the transition to the
ultraviolet (i.e quantum) region may be realized by substitution
of $\frac{\ell^{2}}{\hbar}p_{N_{x^{\mu}}},|N_{x^{\mu}}|\approx 1$
for the quantities $\ell/N_{x^{\mu}}=\frac{\ell^{2}}{\hbar}p_{N_{
x^{\mu}}},|N_{x^{\mu}}|\gg 1$; by the corresponding corrections of
formulae (\ref{Grav1.5})--(\ref{Grav1.5d2}) from Section 3, of all
the components necessary for derivation of gravitational equations
in a {\bf measurable} variant
$\Gamma^\alpha_{\mu\nu}(x,N_{x_{\chi}}),R^{\mu}_{\nu\alpha\beta}(x,N_{x_{\chi}}),...$,
and of formulae (\ref{Grav1.5.3M}),(\ref{Eeq}),... .
\\ Actually, this means that formula (\ref{Grav1.5}) for the
{\it canonically measurable prototype} of the infinitesimal
space-time interval at low energies $E\ll E_p$ is replaced by its
quantum analog  or the {\it canonically measurable quantum
prototype}  for $E\approx E_p$ taking the form
\begin{eqnarray}\label{Grav1.5H} \Delta
s_{\{N_{x_{\chi}}\}}^2(x,{\bf q})\doteq
\frac{\ell^{4}}{\hbar^{2}}g_{\mu\nu}(x,N_{x_{\chi}},{\bf
q})p_{N_{x_{\mu}}}p_{N_{x_\nu}}.
\end{eqnarray}
Here there is no doubt that the numbers $N_{x_\mu},N_{x_\nu}$
belong to the set $\{N_{x_{\chi}}\}$, all the components of this
set are integers with small absolute values, and
$p_{N_{x_{\chi}}}$ are the {\bf generalized measurable} momenta at
high energies corresponding to formulae
(\ref{Deform2Hnew2}),(\ref{Deform2Hnew}).
\\As we have assumed that the values of  $p_{N_{x_{\chi}}}$ are known,
they are {\bf measurable} analogs (or {\bf measurable} variants)
of {\it small space-time variations} at high energies $E\approx
E_p$
\begin{eqnarray}\label{Grav1.5H1}
l_H(p_{N_{x_{\chi}}})\doteq\frac{\ell^{2}}{\hbar}p_{N_{x_{\chi}}};|N_{x_{\chi}}|\approx
1.
\end{eqnarray}
According to formula (\ref{Deform2Hnew2}), we can easily obtain
its variant in the space-time consideration as follows:
\begin{eqnarray}\label{Deform2Hnew2*}
l_H(p_{N_{x_{\chi}}})\stackrel{|N_{x_{\chi}}|\approx 1 \rightarrow
|N_{x_{\chi}}|\gg 1}{\Rightarrow}\frac{\ell}{N_{x_{\chi}}}.
\end{eqnarray}
By the substitution $\ell/N_{x_{\chi}}\mapsto l_H(p_{N_{x_{\chi}}})$
in formulae (\ref{Grav1.5d}),(\ref{Grav1.5d2}) we  can have quantum analogs
of {\it minimal measurable variations} of the metric
and of the partial derivative
\begin{eqnarray}\label{Grav1.5d-H}
\Delta_{\bf q} g_{\mu\nu}(x,N_{x_{\chi}},{\bf q})_{\chi}\doteq
g_{\mu\nu}(x+l_H(p_{N_{x_{\chi}}}),N_{x_{\chi}},{\bf
q})-g_{\mu\nu}(x,N_{x_{\chi}},{\bf q}),
\nonumber\\
\Delta_{\chi,{\bf q}}g_{\mu\nu}(x,N_{x_{\chi}},{\bf
q})\doteq\frac{\Delta_{\bf q} g_{\mu\nu}(x,N_{x_{\chi}},{\bf
q})_{\chi}}{l_H(p_{N_{x_{\chi}}})}.
\end{eqnarray}
Then,using the substitution in formula (\ref{Div1.4new})
\begin{eqnarray}\label{Change-1}
\mathbf{\frac{\ell}{N_{x_\mu}}}\mapsto l_H(p_{N_{x_{\mu}}});
\mathbf{\frac{\Delta}{\Delta_ {N_{x_\mu}}}}\mapsto
\mathbf{\frac{\Delta_q}{\Delta_
{N_{x_\mu},q}}},\nonumber\\\mathbf{\frac{\Delta_q
F(x_\mu)}{\Delta_
{N_{x_\mu},q}}}=\frac{F(x_\mu+l_H(p_{N_{x_{\mu}}}))-F(x_\mu)}{l_H(p_{N_{x_{\mu}}})}
\end{eqnarray}
and applying this substitution to all the formulae in the {\bf
measurable} format of Subsection 5.1, we can derive  at high
energies $E\approx E_p$ all the principal components of Einstein
Equations in the {\bf measurable} form $\mathcal{EEM}$ (formula
(\ref{Eeq-M})). In what follows, we use for them $\bf q$ in
parenthesis, as in formula (\ref{Grav1.5H}).
\\ Substituting these components into Einstein
Equations, we obtain a {\bf measurable} variant of Einstein
Equations at high energies. For short, we denote it as
$\mathcal{EEM}[{\bf q}]$:
\begin{eqnarray}\label{Eeq-M,q}
\mathcal{EEM}[{\bf q}]\doteq R_{\mu\nu}(x,N_{x_{\chi}},{\bf q}) -
\frac12 \,R(x,N_{x_{\chi}},{\bf q})\,
g_{\mu\nu}(x,N_{x_{\chi}},{\bf q})-\nonumber\\ -\frac12 \,
\Lambda(x,N_{x_{\chi}},{\bf q}) \, g_{\mu\nu}(x,N_{x_{\chi}},{\bf
q}) = \nonumber \\ = 8\, \pi \, G\, T_{\mu\nu}(x,N_{x_{\chi}},{\bf
q}).
\end{eqnarray}
As a result, we have
\begin{eqnarray}\label{Eeq-M2,q}
\lim\limits_{E\ll E_p}\mathcal{EEM}[{\bf q}]=\mathcal{EEM} \quad
or \lim\limits_{|N_{x_{\chi}}|\gg 1}\mathcal{EEM}[q]=\mathcal{EEM}.
\end{eqnarray}
For $\mathcal{EEM}[{\bf q}]$ the metrics
$g^{\mu\nu}(x,N_{x_{\chi}},{\bf q})$ (formula (\ref{Grav1.5H})) represent the solutions.
\\
\\{\bf Comment 7.1.}
\\ Thus, at high energies in a {\bf measurable} variant we can have an analog
of Einstein Equations (\ref{Change-1}) and of the metric
(\ref{Grav1.5H}). In the low-energy case $E\ll E_p$ we know the
space-time manifold $\mathcal{M}\subseteq \mathcal{R}^4$ as well
as its minimal variations $\ell/N_{x_\mu}$, whereas at high
energies $E\approx E_p$ in this consideration we, with certainty,
present only ''small'' variations of the space-time positions
$l_H(p_{N_{x_{\chi}}})$ (formula (\ref{Grav1.5H1})).
\\
\\{\bf Comment 7.2.}
\\ When the metric $g_{\mu\nu}(x,N_{x_{\chi}});|N_{x_{\chi}}|\gg 1$
in a {\bf measurable} variant at low energies is varying
practically continuously, the metric
$g_{\mu\nu}(x,N_{x_{\chi}},{\bf q})$; $|N_{x_{\chi}}|\approx 1$
varies discretely and it should experience high fluctuations due
to great fluctuations of the {\bf generalized measurable} momenta
$p_{N_{x_{\chi}}}$, by virtue of $|N_{x_{\chi}}|\approx 1$, in a
good agreement with the results of J. A. Wheeler for the {\it
space-time foam} at the Planck scale
\cite{Wheel},\cite{Wheel1},\cite{mis73},\cite{Ng}--\cite{Scard3}.
\\
\\{\bf Comment 7.3.}
\\ It is obvious that in the {\bf measurable} pattern at high energies
there is no direct analogy with the mathematical apparatus at low
energies (second part of Section 3 and Sections 4,5). Clearly, the
adequate mathematical apparatus in this case should meet the two
following requirements:
\\
\\a)it should be based on the discretely varying quantities expressed
at high energies $E\approx E_p$  in terms  of the {\bf generalized measurable} momenta
$p_{N_{x_{\chi}}};|N_{x_{\chi}}|\approx 1$;
\\
\\b)on going to low energies $E\ll E_p$, i.e. to the region $|N_{x_{\chi}}|\gg 1$,
in accordance with the {\bf Principle of Correspondence},the
above-mentioned mathematical apparatus should present the results
given in Sections 3-5 to a high accuracy.
\\
\\{\bf Comment 7.4.}
\\It is important that Einstein equations for the spherically-symmetric
horizon spaces   \cite{Padm13},derived in the {\bf measurable}
form in \cite{Shalyt-new1},\cite{EJTP} and written at low energies
in terms of the parameter $\alpha_{a}(HUP)\doteq 1/N^{2}_a$ or at
high energies in terms of the parameter $\alpha_{a}(GUP)\doteq
1/[1/4(N_{a}+\sqrt{N_{a}^{2}-1})^{2}]$, where $a=N_{a}\ell$ is a
{\bf primarily} measurable quantity of the space radius,
completely comply with their general form from Sections 5,7.
\\Besides, in accordance with {\bf Remark 2.1}, the condition
$N_{a}\geq 2$ should be fulfilled as noted in {\bf Remark 3.4} of
\cite{EJTP}. This fact was also noted in
\cite{shalyt2},\cite{shalyt3}, however, on the basis of another
approach.

\section{Final Comments and Further Prospects}

In conclusion the author comments on the further course of his studies an the problems involved.
\\
\\{\bf 8.1.} As at low energies $E\ll E_p$ all the spatial variables
$\{x_i\},i=1,2,3$ are equitable, all the numbers
$N_{x_i}\equiv N_{\Delta x_i},
|N_{x_i}|\gg 1$ should be sufficiently close.
\\ At high energies $E\ll E_p$ the numbers $N_{x_i}$ should meet the condition $|N_{x_i}|\approx 1$
and it seems that they are close from the start. But in this case
there are two objections:
\\a) at high energies the essence of the notions ''close'' and ''far from'' may be different;
\\b) in some currently used models it is assumed that at high
energies the space becomes noncommutative (for example,
\cite{Kempf},\cite{Nozari},\cite{Tawf})) and in this case the
spatial positions could cease to be equitable.
\\
\\{\bf 8.2.} According to Remark {\bf 1.d)}, $\ell$ -- minimal {\bf
primarily measurable} length. At the beginning of this paper, we
have assumed that it is close to $l_p$, i.e. $\ell\propto l_p$ or
same $\ell=\kappa l_p,\kappa\approx1$. But, actually, all the
results of this paper, except of those associated with the
simplest form of GUP (\ref{U2}), are independent of the magnitude
of the {\bf primarily measurable} length $\ell$. When $\ell\propto
l_p\approx 10^{-33}cm$, it is clear that for $|N|\gg 1$ lengths of
the form $\ell/N$ are very small.
\\ Of course, for experimental physics at high energies far from the Planck energies,
e.g. for those in LHC, $\ell$ may exceed the Planck length
$\ell\gg l_p$ considerably and in this case all the fundamental
energies available in  LHC satisfy the condition  $E\ll E_\ell\ll
E_p$. Then $\ell$ (and hence $E_\ell$) determines the natural
ultraviolet-cutoff bound in the corresponding quantum theory.
because of this, the following problem arises.
\\
\\{\bf 8.3.} A correct (without ultraviolet and infra-red divergences)
quantum theory with the parameter $\ell$ (same $E_\ell$) should be
resolved in conformity to all experimental data of LHC. The
behavior of such a theory in the high-energy (ultraviolet) region
should be associated with the quantity $\ell$. The absence of
infra-red divergences should be given by the natural upper bound
for $N_{x_i}$ and determined from formulae
(\ref{MOD-1})--(\ref{Semic}) in Section 6 that, due to the
condition $\ell\gg l_p$, would be significantly lower and hence
more realistic.
\\ Moreover, for $\ell\approx l_p$, due to infinitesimal values of
$\ell/N;|N|\gg 1$, the corresponding ''{\bf measurable}'' theory,
still being discrete, would be very close to (practically
indistinguishable from) the initial continuous theory enabling us
to solve the problems mentioned in points {\bf 6.1.} and {\bf
6.2.}. But for $\ell\gg l_p$, considering formulae
(\ref{MOD-1})--(\ref{Semic}), the situation may be different.
\\ Indeed, as points {\bf 6.1.} and {\bf 6.2.} conform to the
semiclassical approximation, i.e. to consideration of the
quantized matter fields against the classical space-time pattern,
the {\bf primarily measurable} minimal length $\ell\gg l_p$ is
more adequate to study these problems.
\\ It seems that the absence of infra-red divergences directly
follows from the lower limit for the momenta in a quantum theory
with different generalizations of the Uncertainty Principle (for
example, \cite{Tawf}).
\\
\\{\bf 8.4.} So, in this paper it is proposed in a quantum theory
and in gravity to use, instead of abstract small and infinitesimal
quantities $\delta x_\mu,dx_\mu,...$, the small quantities derived
from the {\bf primarily measurable} minimal length $\ell$ with the
help of the corresponding formulae in Sections 2-6 of the paper.
Note that $\ell$ may be both close to the Planck length
$\ell\propto l_p$ and considerably higher than this length
$\ell\gg l_p.$
\\In the suggested paradigm all variations of a physical system which may be regarded small,
as distinct from the continuous space-time consideration, have
particular values determined at low energies $E\ll E_\ell$ by the
{\bf primarily measurable} momenta and at high energies $E\approx
 E_\ell$ by the {\bf generalized measurable} momenta from Section 2.
In this way all small variations are dependent on the existing
energies.
\\Since at the present time no direct or indirect experiments
at the scales on the order of Planck’s scales (i.e. at the
energies associated with the quantum gravity scales) are known,
all theoretical studies in this field are to some or other extent
speculative.  Nevertheless, considering that gravity should be
formulated with the use of the same terms at all the energy
scales, it must be governed by the particular unified principles
the formulation of which varies depending on the “available”
energies. Because of this, the results from Section 7 seem to be
important. Of course, these results are tentative and may be
corrected during further studies of gravity in terms of the {\bf
measurability} notion. But they give the main idea and define the
trend towards the derivation of a {\bf measurable} variant of
gravity: framing of a correct gravitational theory at all the
energy scales, with the use of a set of discrete parameters
$p(N_{\Delta x_\mu})$ for all nonzero integer values of $N_{\Delta
x_\mu}$, that is close to the General Relativity at low energies
$E\ll E_p$ and is a new (discrete) theory at high energies
$E\approx E_p$.
\\
{\bf 8.5.} In conclusion it may be stated that the principal
result of this work is as follows.
\\{\bf 8.5.1.} At low energies far from the Planck energies $E\ll E_p$
we replace the space-time manifold $\mathcal{M}\subseteq
\mathcal{R}^4$ by the lattice model (denoted by
$Latt^{LE}_{\{N_{x_\mu}\}}\mathcal{M}$, where the upper index
$^{LE}$ is the abbreviation for ''Low Energies''), with the nodes
taken at the points $\{x_\mu\}\in \mathcal{M}$ so that all the
edges belonging to $\{x_\mu\}$ have the size
$\ell/N_{x_\mu}$,where $N_{x_\mu}$ - integers having the property
$|N_{x_\mu}|\gg 1.$ As the edge lengths $\ell/N_{x_\mu}$, within a
constant factor, are coincident with  the {\bf primarily
measurable} momenta (formula (\ref{Meas-D4})),the  model
$Latt^{LE}_{\{N_{x_\mu}\}}\mathcal{M}$ is dynamic and dependent on
the existing energies. In this case all the main attributes of a
gravitational theory in the manifold $\mathcal{M}$,including
Einstein Equations, have their adequate analogs on the
above-mentioned lattice $Latt^{LE}_{\{N_{x_\mu}\}}\mathcal{M}$,
giving the low-energy deformation of General Relativity in terms
of paper \cite{Fadd} (Section 4).
\\
\\{\bf 8.5.2.} At high Planck’s energies $E\propto E_p$,
the lattice model $Latt^{LE}_{\{N_{x_\mu}\}}\mathcal{M}$ is
replaced by $Latt^{HE}_{\{N_{x_\mu}\}}\mathcal{M}$ (the upper
index $^{HE}$ is the abbreviation for ''High Energies''), the
edges with the lengths $\ell/N_{x_\mu}$ are replaced by those with
the lengths $l_H(p_{N_{x_{\mu}}})$ which, within a constant
factor, are coincident with the {\bf generalized measurable}
momenta $p_{N_{x_{\mu}}}$,where $N_{x_{\mu}}$-integer number
having the property $|N_{x_{\mu}}|\approx 1$ (formula
(\ref{Grav1.5H1})). In this way
$Latt^{HE}_{\{N_{x_\mu}\}}\mathcal{M}$ also represents a dynamic
model that is dependent on the existing energies and may be the
basis for the construction of a correct variant of the high-energy
deformation in General Relativity (Section 5).

\begin{center}
{\bf Conflict of Interests}
\end{center}
The author declares that there is no conflict of interests
regarding the publication of this paper.

\end{document}